# A Probabilistic Framework for Predicting Spatiotemporal Intensity and Variability of Outdoor Thermal Comfort

Shisheng Chen[a, b], Ruohan Xu[b, *], Nyuk Hien Wong[b], Shanshan Tong[b], Jiashuo Wang[c], Matthaios Santamouris[c]

a-School of Architecture and Urban-rural Planning, Fuzhou University, P.R. China 350108.

b-Department of the Built Environment, National University of Singapore, Singapore 117566.

c-School of Built Environment, University of New South Wales, Australia 2052.

*Corresponding author.

E-mail address: bdgxr@nus.edu.sg (R. Xu)

## Abstract

Thermal conditions in the urban canopy exhibit stochastic variability driven by varied radiative fluxes and turbulent wind fields requiring probabilistic rather than deterministic prediction methods. This study presents a probabilistic framework for predicting the spatial and temporal intensity and variability of outdoor thermal comfort in tropical urban environments. The framework integrated ground measured meteorological data and remote sensing urban morphological data to calculate Physiological Equivalent Temperature (PET), and uses K-means, XGBoost and Monte Carlo simulations on PET's training and inference. The prediction model achieved strong predictive performance, with $R^2$, RMSE, and SMAPE values of 0.93, 0.81 °C, and 1.34% for $PET_{mean}$, and 0.85, 0.38 °C, and 10.44% for $PET_{std}$, respectively. Case study showed clear spatial heterogeneity of outdoor thermal comfort. Locations with dense tree canopies and vegetated surfaces displayed the normalized percentage of acceptable thermal comfort (NATC) up to 65%, whereas built-up zones dominated by impervious surfaces, such as industrial estates and high-density residential areas, recorded NATC below 30%. Greenery was found to mitigate both the intensity of heat stress and its variability, producing a stable and comfortable microclimate. Daytime $PET_{std}$ ranged from 4.0–4.5 °C in built-up areas to 1.5–2.0 °C in greenery-covered zones, while nighttime $PET_{std}$ decreased to 2.2–2.4 °C and 1.2–1.4 °C, respectively. These findings emphasize the critical role of greenery in mitigating thermal variability and enhancing outdoor thermal comfort, while revealing the stochastic nature of thermal comfort across different urban morphologies.

**Keywords**: Outdoor Thermal Comfort; Physiological Equivalent Temperature; Heat Stress; Clustering; Regression; Probabilistic Prediction

## Nomenclature

| Symbol | Meaning |
| --- | --- |
| PET | Physiological equivalent temperature |
| QUEST | Quantitative urban environment simulation tool |
| GEE | Google earth engine |
| PCA | Principal component analysis |
| SC | Silhouette coefficient |
| DBI | Davies–bouldin index |
| CHI | Calinski–harabasz index |
| SMAPE | Symmetric mean absolute percentage error, % |
| RMSE | Root mean squared error, °C |
| $R^2$ | Coefficient of determination |
| $PET_{mean}$ | Mean of physiological equivalent temperature, °C |
| $PET_{std}$ | Standard deviation of physiological equivalent temperature, °C |
| $\mathrm{T_a}$ | Dry-bulb air temperature, °C |
| $\mathrm{T_{mrt}}$ | Mean radiant temperature, °C |
| $\mathrm{V_a}$ | Air speed, m/s |
| RH | Relative humidity, % |
| NDVI | Normalized difference vegetation index |
| MNDWI | Modified normalized difference water index |
| NDBI | Normalized Difference Built-up Index |
| BSI | Bare soil index |
| $\alpha_{land}$ | Land surface albedo |
| $\epsilon_{land}$ | Land surface emissivity |
| *NIR* | Near-infrared reflectance |
| *SWIR* | Shortwave infrared reflectance |
| $Pv$ | Fractional vegetation cover |
| $\epsilon_s$ | Soil emissivity |
| $\epsilon_v$ | Vegetation emissivity |
| NATC | Normalized percentage of acceptable thermal comfort, % |
| $N_{c,h}$ | Number of observations belonging to acceptable thermal comfort at hour $h$ |
| $N$ | Total number of observations over at hour $h$ |
| $c$ | Set of acceptable heat stress categories including slight cold stress, no heat stress, slight heat stress |
| $\text{Rain}_{\text{tot}}$ | Total rainfall, mm |
| $\text{Wind}_{\text{avg}}$ | Wind speed, m/s |
| $\text{Solar}_{\text{tot}}$ | Total global horizontal solar irradiance, $W/m^2$ |
| $\text{Solar}_{\text{max}}$ | Maximum global horizontal solar irradiance, $W/m^2$ |
| $\text{Solar}_{\text{avg}}$ | Mean global horizontal solar irradiance, $W/m^2$ |
| $T_{\text{avg}}$ | Average air temperature, °C |
| $T_{\text{max}}$ | Maximum air temperature, °C |
| $T_{\text{min}}$ | Minimum air temperature, °C |
| $Tair_{ref}$ | Reference dry-bulb air temperature, °C |
| $Solar_{ref}$ | Reference global horizontal solar irradiance, $W/m^2$ |
| $V_{a_{ref}}$ | Reference wind speed, m/s |

# 1. Introduction

## 1.1 Outdoor Thermal Comfort Assessment

Assessing outdoor thermal comfort is essential for enhancing urban heat resilience and can be approached through objective assessments and subjective questionnaires. Objective assessments typically rely on either microclimate simulations or onsite measurements. Simulations such as Computational Fluid Dynamics (CFD) are suitable for large-scale evaluations but require extensive computational resources and validation. In contrast, onsite measurements offer higher spatial accuracy but are limited by the number of sampling points. Both approaches face challenges in accuracy due to the need to interpolate data across unmeasured areas, which introduces potential uncertainty into thermal comfort assessments.

Several integrative thermal indices derived from the human energy balance such as physiological equivalent temperature (PET) (Höppe, 1999), universal thermal climate index (UTCI) (Bröde et al., 2012), and outdoor standard effective temperature (OUT_SET*) (Spagnolo and De Dear, 2003) have been developed to quantify human thermal comfort.

The Physiological Equivalent Temperature (PET) is an important index for evaluating human thermal comfort in outdoor environments, particularly in the context of urban heat islands and occupational health and safety (Farhadi et al., 2019; Fiorillo et al., 2023; Hwang et al., 2023). PET is derived from the Munich Energy Balance Model for Individuals (MEMI) by Hoppe and represents the air temperature at which the human energy budget is maintained under specific conditions of skin temperature and core temperature (Höppe, 1999). PET enables a comparison of complex thermal conditions outside with one's own experience indoors, allowing for a more intuitive understanding of thermal comfort or discomfort. Walther and Goestchel propose a correction to the PET calculation defined by Hoppe and compare the corrected model with the original method, highlighting errors in calculation routine and vapor diffusion model (Walther and Goestchel, 2018). For example, the original PET model uses a simplified approach to solve the equation system, while the corrected model provides a more rigorous solution, leading to more accurate assessments of thermal comfort. In addition, the vapor diffusion model of the original PET model is not dependent on the clothing insulation, an issue addressed by the corrected model by aligning it with state-of-the-art vapor transport models. The corrected PET model of Walther and Goestchel is later adopted by the pythermalcomfort package using dry-bulb air temperature ($T_a$), mean radiant temperature ($T_{mrt}$), air speed ($V_a$), relative humidity (RH), metabolic rate and clothing insulation as input parameters to calculate PET (Tartarini and Schiavon, 2020).

Universal Thermal Climate Index (UTCI) was developed by International Society of Biometeorology (ISB) to provide a thermal comfort model suitable for outdoor environment (Błażejczyk et al., 2010). UTCI evaluates outdoor thermal conditions using a single metric, reflecting human physiological responses to the multidimensional outdoor environment including air temperature, wind speed, humidity, and longwave and shortwave radiation (Bröde et al., 2012). Human responses are simulated using the UTCI-Fiala multi-node thermoregulation model coupled with an adaptive clothing model to account for behavioural and physiological adjustments (Fiala et al., 2012).

The outdoor standard effective temperature (OUT_SET*) is an extension of indoor standard effective temperature (SET*) adapted for outdoor conditions. It accounts for air and mean radiant temperatures, relative humidity, air velocity, clothing insulation, and activity level (Spagnolo and De Dear, 2003). Its actual mean radiant temperature is computed using a human thermoregulation model. Compared with PET and UTCI, OUT_SET* is less commonly applied, and its predictive performance remains to be fully evaluated.

At neighbourhood scales, urban land cover and morphology systematically modulate the radiative and convective exchanges that shape local microclimate and outdoor thermal comfort. Recent studies increasingly use remotely sensed indicators as standard proxies for these controls. For example, vegetation indices capture shading and evapotranspiration cooling (Carlson, 2019; Chen et al., 2006; Dobrovolný and Krahula, 2015; Imran et al., 2021; Jamei et al., 2022; Jia and Wang, 2020), and water-related indices reflect advection and the thermal inertia of open water (Chen et al., 2006; Imran et al., 2021; Jia and Wang, 2020; Xu, 2006). In addition, built-up and bare soil metrics indicate imperviousness and surface moisture status (Chen et al., 2020, 2020, 2006; Guo et al., 2022; Imran et al., 2021; Jamei et al., 2022; Jia and Wang, 2020; Zha et al., 2003), and broadband albedo and surface emissivity parameterize short-wave reflection and long-wave exchange (Chen et al., 2023; He et al., 2025; Musco et al., 2016; Tabatabaei and Fayaz, 2023; Vuckovic et al., 2016). Consistent empirical evidence shows that higher fractions of vegetation and water are associated with reduced surface thermal loads and improved outdoor conditions, whereas greater imperviousness and exposed soils intensify heat exposure (Chatterjee and Majumdar, 2022; Imran et al., 2021; Jamei et al., 2022; Kumar Gavsker, 2023). These findings motivate the use of satellite-derived morphology as explanatory covariates for spatially explicit thermal comfort assessments when paired with ground-level meteorology.

### 1.2 The Sampling Framework and Analysis of Outdoor Thermal Comfort

PET assessment procedures involve a sequence of microclimate monitoring, PET calculation, heat stress classification, and impact evaluation. Firstly, PET is derived from high-resolution microclimate measurements, including 1-min logging of $T_a$, RH, $V_a$, and $T_{globe}$ at various heights such as pedestrian level, mid-level, and rooftop (Miao et al., 2023). Similar data were collected through 1-min observations at twelve factory sites during three clear winter days in Haining (He et al., 2023), 1-min to 30-min mobile scans at 23 street canyon points in Harbin from July 2020 to January 2021 (Sun et al., 2022), and 1-min to 5-min logging within two sunken squares during two seven-day surveys in winter and summer in Shenzhen (Zhao et al., 2023).

PET is then computed using tools like RayMan (He et al., 2023; Miao et al., 2023) or ENVI-met BIO-met, incorporating measured environmental parameters along with clothing insulation and metabolic rates derived from questionnaire responses (Tabatabaei and Fayaz, 2023). Thermal sensation votes are then correlated with PET values to calculate the neutral PET ranges or 80% acceptability ranges. For instance, the correlation between mean thermal sensation votes with PET binned at 1 °C intervals yields neutral PET ranges of 14.3 °C in winter (He et al., 2023), 29 °C in summer (Zhao et al., 2023), and 14.5–22.8 °C across four seasons (Sun et al., 2022), as well as 80% acceptability ranges of 7.9–27.3 °C at cold region (Sun et al., 2022) and 10.7–17.8 °C in humid subtropical climate (He et al., 2023). In addition, regression analysis is performed to quantify the influence of urban design factors on PET, such as façade material properties (e.g., solar reflectance, emissivity) (Lopez-Cabeza et al., 2022; Tabatabaei and Fayaz, 2023), and street geometry variables (e.g., sky view factor, aspect ratio) (Muniz-Gäal et al., 2020; Sun et al., 2022). These field measurements are frequently supplemented by simulations. For instance, hourly TMY3-driven Ladybug modelling was conducted for 18 shading configurations over three summer months in Taichung to validate the PET condition (Ou and Lin, 2023). The impact of twenty façade materials on PET was evaluated through 24-hour ENVI-met BIO-met running on summer and winter days in Shiraz (Tabatabaei and Fayaz, 2023).

Translating morphology and meteorology into site-specific and hour-by-hour comfort estimates benefits from data-driven models that can learn nonlinear interactions while remaining interpretable (Lamberti et al., 2023). Predictive models coupled with machine learning interpretable techniques such as SHAP analysis meet this need as the models can estimate not only the expected value of PET but also its dispersion (e.g., standard deviation), and SHAP provides an additive and physically interpretable attribution of each predictor's contribution (Li et al., 2025). Together, these tools enable probabilistic and explainable mapping of outdoor thermal comfort.

### 1.3 Research Gap and Objectives

Microclimatic conditions within the urban canopy are governed by varied radiation fluxes and complex wind fields leading to significant spatiotemporal variability in thermal comfort perception. This stochastic variability arises from the interactions between surface-atmosphere energy exchange, turbulence, and localized shading effects, and therefore cannot be reliably represented by deterministic methods. Instead, probabilistic modelling tools are needed to capture the likelihood distribution of thermal comfort outcomes. Although PET is frequently used in thermal comfort assessments, existing studies exhibit several limitations. Most studies rely on deterministic approaches employing fixed values or averaged environmental conditions for PET calculation, which neglects the inherent variability and uncertainty in meteorological and human-related parameters. In addition, many surveys are based on short monitoring periods and fail to capture the large fluctuations in temperature, humidity and wind speed that can occur even under similar hot weather conditions when observed over longer periods. These limitations highlight the need for approaches accounting for variability and uncertainty when evaluating thermal comfort. In this study, a probabilistic framework incorporating variability statistics (e.g., standard deviation) and stochastic modelling (e.g., Monte Carlo simulation) into outdoor thermal comfort modelling has been developed under clear and hot weather conditions. The framework not only predicts the expected mean PET but also quantifies its uncertainty through the distribution of PET values. The objectives of this study were to: (a) evaluate spatial and temporal patterns of intensity and variability of PET across an urban region; (b) identify key urban morphological features influencing both the mean and standard deviation of PET; (c) explain the morphological drivers of thermal hotspots to support targeted heat mitigation strategies.

The remaining structure of the paper is organized as follows. The research design is described in Section 2, including data collection of meteorological data, remote sensing data, calculation of outdoor thermal comfort, and probabilistic prediction framework. Section 3 presents the results of spatial and temporal heterogeneity of PET across sampled locations and prediction performance of probabilistic models. Section 4 and Section 5 discuss and concludes the potential and limitation of probabilistic models and key drivers behind the intensity and variability of outdoor thermal comfort.

## 2. Methodology

### 2.1 Research Design

The research framework (Figure 1) integrates ground-measured meteorological observations with multispectral remote sensing data to characterize the spatiotemporal heterogeneity of urban thermal comfort. Following rigorous quality control, microclimatic data were employed to derive hourly PET values across sampled locations. K-means clustering was applied to classify reference meteorological data conditioned by radiation, temperature, and wind. XGBoost regression was applied to link PET with environmental and morphological predictors at unsampled locations. The Monte Carlo simulation based on trained models produced continuous spatial mapping of PET and NATC, providing high-resolution maps of heat exposure to support planning and policy assessment.

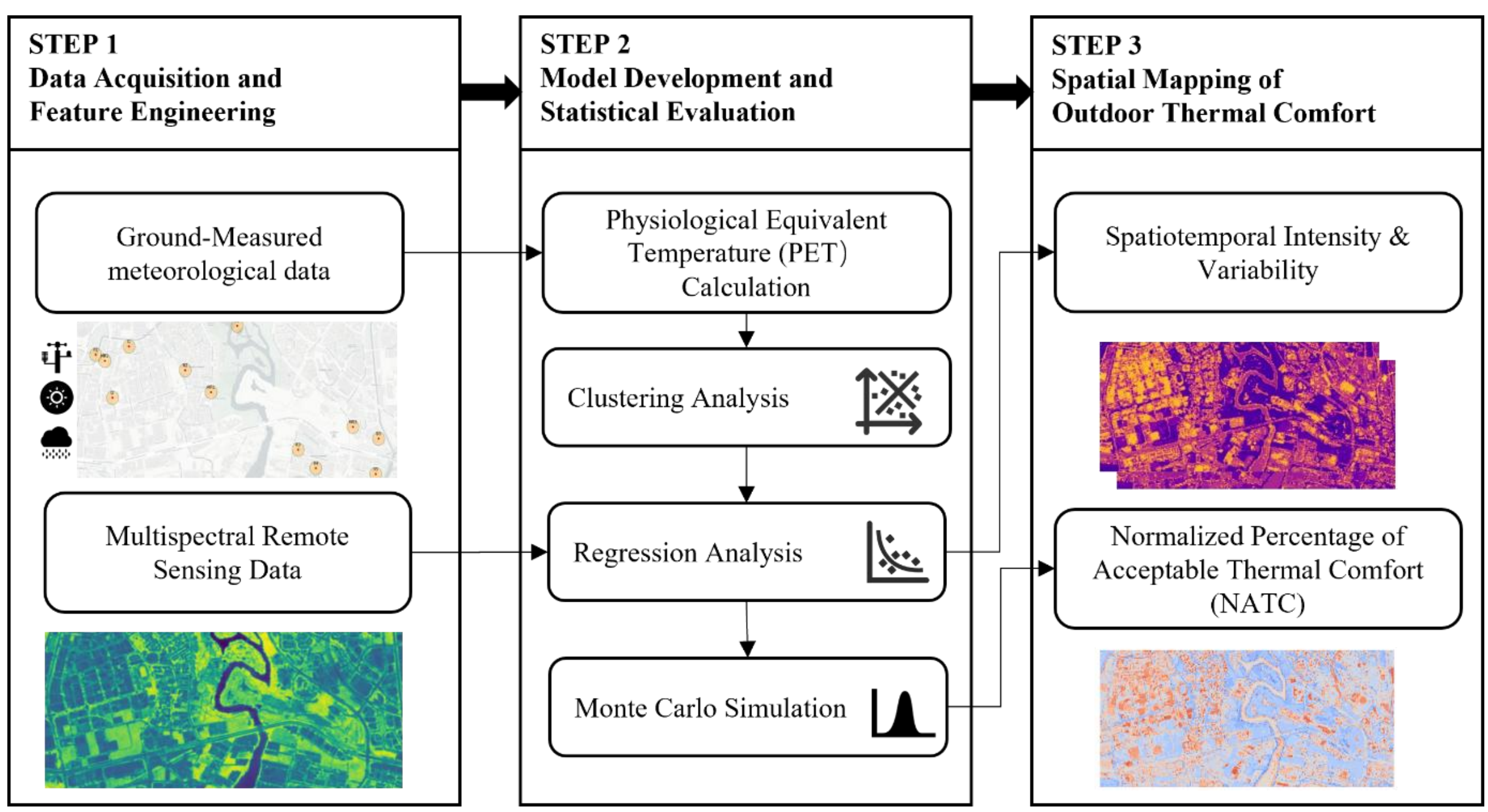

Figure 1: Probabilistic prediction framework for outdoor thermal comfort

### 2.2 Collection of Meteorological Data

This study demonstrated the prediction performance of proposed model using the ground-level microclimate data collected from Quantitative Urban Environment Simulation Tool (QUEST), a platform for analysing and testing the immediate microclimatic impact of development plans and assessing their long-term impacts under future climate change scenarios in Singapore (Lim et al., 2021). The QUEST data were collected in the eastern part of Singapore, recording meteorological data including $T_{globe}$, $T_a$, RH, and $V_a$ at 5-minute interval from Feb 19th 2018 to Dec 8th 2018. The selected measurement points are shown in the OpenStreetMap (Figure 2) (Boeing, 2024). Points prefixed with R and B indicate measurement locations in residential areas and business areas, respectively. Points prefixed with I and P indicate measurement locations in industrial areas and parks, respectively. Points prefixed with MR represent major roads.

Table 1 illustrated the instruments used in this study. $T_a$ and RH were measured using ONSET U23-001 HOBO PRO V2 INT TEMP/RH. $V_a$ was measured using ONSET S-WSB & WDA -M003. $T_{globe}$ was measured using ONSET UX100-014M T DATA LOGGER with grey ping pong and thermo-couple wire.

$T_{mrt}$ was calculated from Equation (1) based on the calibrated mean convection coefficient using a 40 mm grey globe with an emissivity of 0.97 in Singapore (Tan et al., 2013). The calibrated $T_{mrt}$ equation is only valid for $0\ \mathrm{m/s} \leq V_a \leq 4\ \mathrm{m/s}$ (Tan et al., 2013) used as the criterion for filtering out meteorological data.

$$T_{mrt} = \left[\left(T_{globe} + 273.15\right)^4 + \frac{3.42 \times 10^9 V_a^{0.119}}{\varepsilon \times D^{0.4}} \times \left(T_{globe} - T_a\right)\right]^{0.25} - 273.15 \tag{1}$$

Where $T_{globe}$ is the globe temperature (°C); $T_a$ is the air temperature (°C); $V_a$ is the wind speed (m/s); D is the globe diameter (mm); ε is the emissivity of globe.

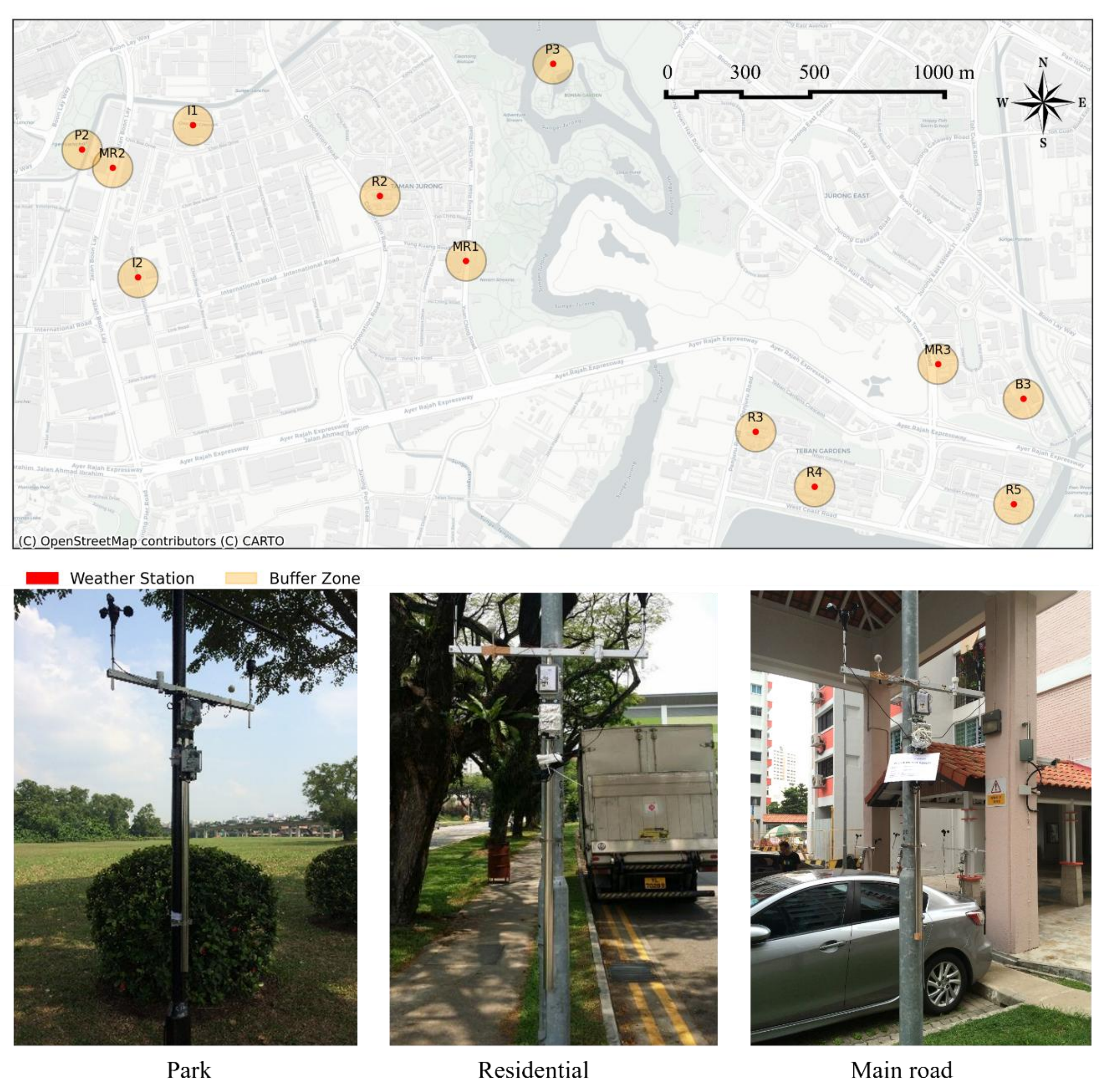

Figure 2: Locations of measurement points

Table 1: Specifications of instruments

| Sensors | Variable(s) | Measurement range | Accuracy | Sampling rate |
|---|---|---|---|---|

| ONSET U23-001 HOBO PRO V2 INT TEMP/RH | $T_a$ & RH | -40° to 70°C<br>0-100% | ±0.21°C<br>±2.5% | 5-minute |
|---|---|---|---|---|
| ONSET S-WSB | $V_a$ | 0 to 76 m/s | ±1.1m/s | 5-minute |
| ONSET UX100-014M T DATA LOGGER with grey ping pong and thermo-couple wire | $T_{globe}$ | -20° to 70°C | ±0.21°C | 5-minute |

**2.3 Collection of Multispectral Remote Sensing Data**

To characterize the land surface conditions surrounding each sampling point, a group of remote sensing indices were derived using Sentinel-2 satellite multispectral data (surface reflectance imagery) processed via Google Earth Engine (GEE) (Gorelick et al., 2017). The imagery was filtered for cloud cover (<20%) within the study period (2018–2020) and spatially constrained to a specified buffer radius of 100m around each point. For each sampling point, several indices representing vegetation, water, and built-up area were calculated including the Normalized Difference Vegetation Index (*NDVI*), Modified Normalized Difference Water Index (*MNDWI*), Normalized Difference Built-up Index (*NDBI*), Bare Soil Index (*BSI*), land surface albedo ($\alpha_{land}$), and land surface emissivity ($\epsilon_{land}$) using corresponding Sentinel-2 satellite bands, as shown in Table 2. The surface reflectance imagery was used to extract mean values for each index within the buffer zone. These remotely sensed metrics were subsequently used as independent variables in the machine learning models for predicting outdoor thermal comfort conditions. The computed urban morphological values are shown in Figure 3.

Table 2: Selected Sentinel-2 satellite bands used in this study

| Band | Central Wavelength (nm) | Resolution (m) | Band Number (Sentinel-2) |
|---|---|---|---|
| Blue | 490 | 10 | B2 |
| Green | 560 | 10 | B3 |
| Red | 665 | 10 | B4 |
| NIR | 842 | 10 | B8 |
| SWIR1 | 1610 | 20 | B11 |
| SWIR2 | 2190 | 20 | B12 |

*NDVI* is based on the difference of reflectance in the near-infrared and red bands (Lillesand et al., 2015):

$$NDVI = \frac{NIR - Red}{NIR + Red} \tag{2}$$

where *NIR* is the near-infrared reflectance and *Red* is the red reflectance.

*NDBI* is referred to the difference of reflectance between TM Band 4 (0.76-0.9μm) and TM Band 5 (1.55-1.75μm) bands (Zha et al., 2003):

$$NDBI = \frac{SWIR1 - NIR}{SWIR1 + NIR} \tag{3}$$

where *SWIR1* is the shortwave infrared reflectance with a central wavelength of 1.61μm.

*MNDWI* is originally derived based on TM Band 5 (1.55-1.75μm) and TM Band 2 (Green) from Landsat 5 (Xu, 2006):-

$$MNDWI = \frac{Green - SWIR1}{Green + SWIR1} \tag{4}$$

where *Green* is the green reflectance.

*BSI* represents exposed soil surfaces and uncultivated areas and is formulated using Equation (Mzid et al., 2021; Rikimaru et al., 2002):

$$BSI = \frac{(SWIR1 + Red) - (NIR + Blue)}{(SWIR1 + Red) + (NIR + Blue)} \tag{5}$$

where *Blue* is the blue reflectance.

$\alpha_{land}$ is estimated as a weighted combination of visible, NIR, and shortwave infrared reflectance bands (Smith, 2010):

$$\alpha_{land} = \frac{0.356Blue + 0.130Red + 0.373NIR + 0.085SWIR1 + 0.072SWIR2 - 0.0018}{1.016} \tag{6}$$

where *SWIR2* corresponds to Sentinel-2 Band 12.

$\epsilon_{land}$ was estimated from NDVI based on fractional vegetation cover approach (Carlson, 2019; Sobrino et al., 2008, 2004):

$$Pv = \left(\frac{NDVI - 0.2}{0.5 - 0.2}\right)^2 \tag{7}$$

$$\epsilon_{land} = Pv \times \epsilon_v + (1 - Pv) \times \epsilon_s + 0.005 \tag{8}$$

Where $Pv$ is the fractional vegetation cover; $\epsilon_v$ is the vegetation emissivity with field measured value of 0.985±0.007 (Sobrino et al., 2008; Valor and Caselles, 1996); $\epsilon_s$ is the soil emissivity with field measured value of 0.96±0.01 (Valor and Caselles, 1996).

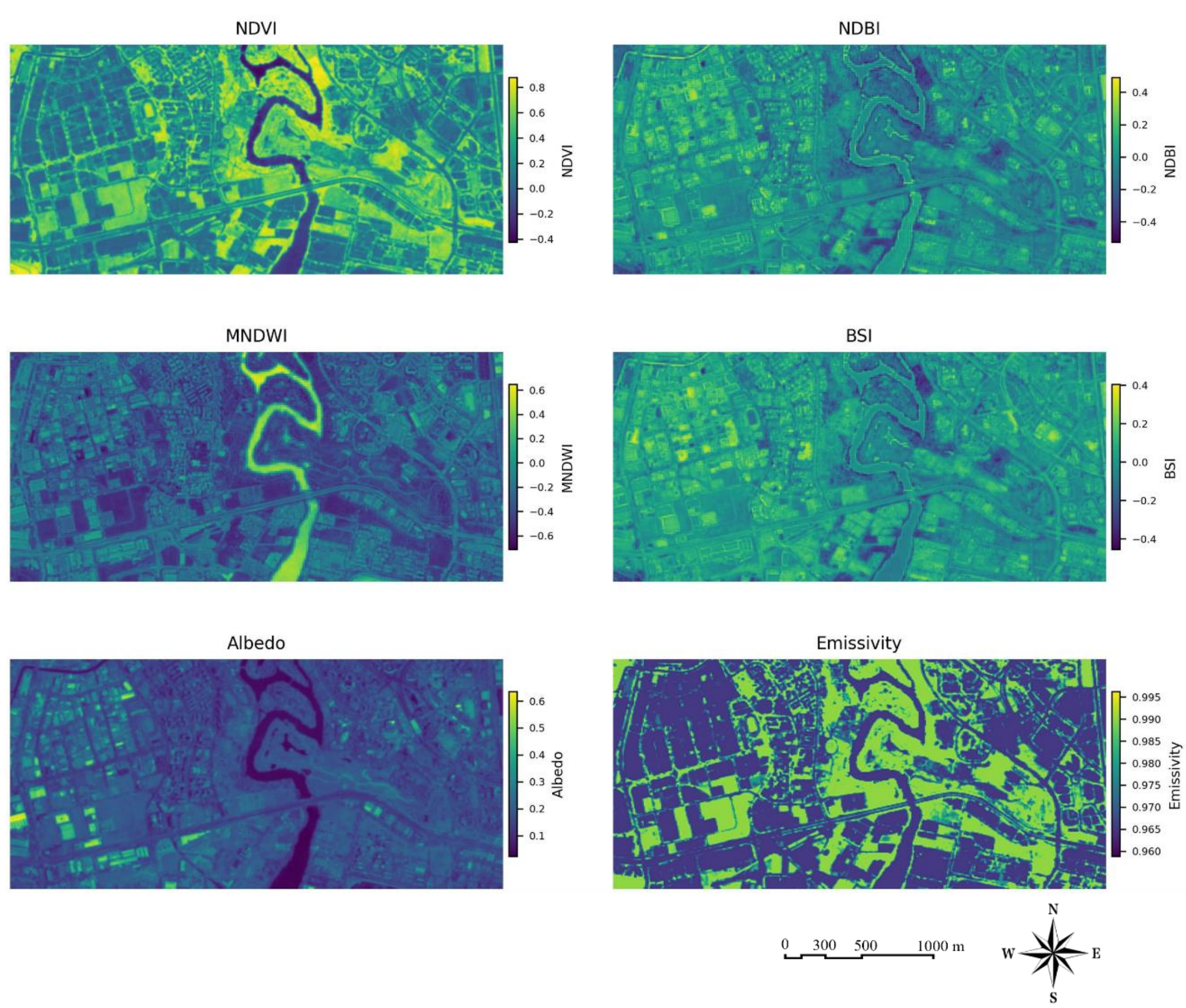

Figure 3: Urban morphological index extracted from satellite images

## 2.4 Calculation of Outdoor Thermal Comfort

Summer metabolic rate and clothing insulation levels were set to 2.0 met and 0.5 clo, respectively, in accordance with ASHRAE Standard 55 (ASHRAE, 2017), representing individuals walking at 0.9 m/s while wearing typical summer attire. PET calculation model was running through pythermalcomfort (Tartarini and Schiavon, 2020) with input features including $T_a$, $T_{mrt}$, $V_a$, and RH. Each simulated PET value was subsequently classified into heat stress levels based on thresholds (Figure 4) calibrated by Singapore's Building and Construction Authority based on thermal comfort perception studies involving local residents (Building and Construction Authority, 2021).

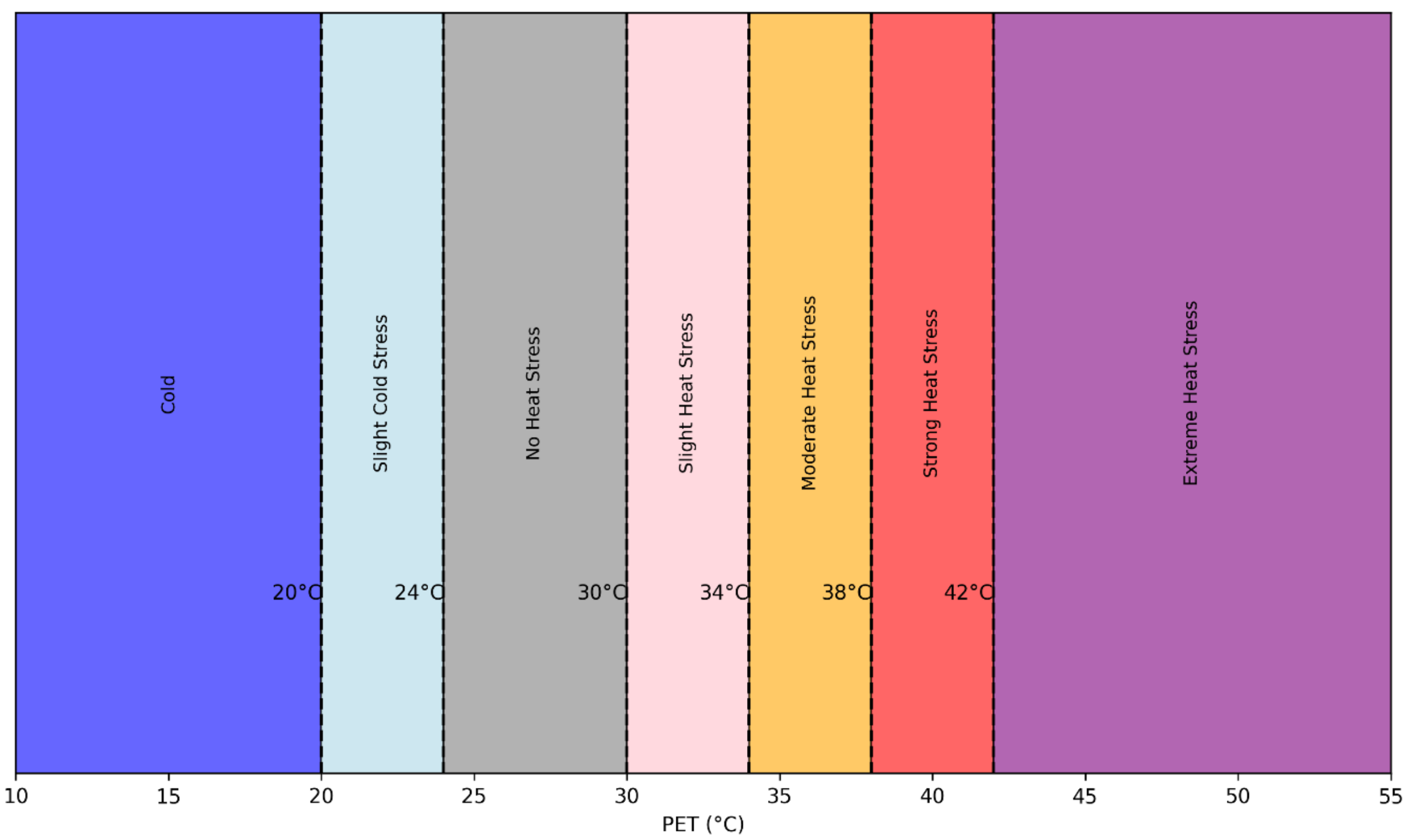

Figure 4: Heat stress level classification of PET used in Singapore

This study considered PET values below 34°C excluding moderate to extreme heat stress as acceptable thermal comfort suitable for outdoor activities. Normalized percentage of acceptable thermal comfort (NATC) is then formulated:

$$NATC(\%) = \frac{1}{24}\sum_{h=0}^{23}\left(\frac{N_{c,h}}{N} * 100\%\right) \tag{9}$$

Where $N_{c,h}$ is the number of observations belonging to acceptable thermal comfort at hour $h$; $N$ is the total number of observations over at hour $h$; $c$ is the set of acceptable heat stress categories (Slight Cold Stress, No Heat Stress, Slight Heat Stress).

## 2.5 Probabilistic Prediction Framework

### *2.5.1 Clustering of Reference Meteorological Data*

The cluster analysis was initiated with reference meteorological data as inputs. These hourly data were aggregated into daily metrics consisting of total rainfall ($Rain_{tot}$), average wind speed ($Wind_{avg}$), total global horizontal solar irradiance ($Solar_{tot}$), maximum global horizontal solar irradiance ($Solar_{max}$), mean global horizontal solar irradiance ($Solar_{avg}$), air temperature ($T_{avg}$), maximum air temperature ($T_{max}$), and minimum air temperature ($T_{min}$).

The resulting daily data were then normalized to a 0–1 range before clustering. The K-means algorithm implemented using sklearn.cluster.KMeans (Arthur and Vassilvitskii, 2007; Pedregosa et al., 2011) was employed to group the samples by minimizing inertia. To enhance the clustering performance, Principal

Component Analysis (PCA) was applied, projecting the original data into a lower-dimensional space. In this study, PCA was implemented using sklearn.decomposition.PCA (Pedregosa et al., 2011), with the number of components set to 2 thereby reducing the dataset to a two-dimensional representation. Clustering was performed with the number of clusters ranging from 2 to 7.

Inertia measures the compactness of clusters by summing squared distances between each point and its cluster centroid:

$$Inertia = \sum_{j=1}^{k} \sum_{i=1}^{n} min\left\| x_i^j - c_j \right\|^2 \tag{10}$$

Where k is the number of clusters, n is the number of samples, $x_i^j$ is sample i in cluster j, $c_j$ is centroid of cluster j. After running the clustering algorithm, each sample of meteorological data was labelled by a numerical value to indicate the classification of clusters (e.g. 0 for cluster 0). Lower inertia indicates more compact clusters.

In addition, this study also applied Silhouette Coefficient (SC), Davies–Bouldin Index (DBI), Calinski–Harabasz Index (CHI) from sklearn.metrics (Pedregosa et al., 2011) for clustering evaluation.

SC measures both intra-cluster distance and nearest-cluster distance. For each sample:

$$\text{SC (i)} = \frac{b(i) - a(i)}{\max\left(a(i) - b(i)\right)} \tag{11}$$

Where a is the mean intra-cluster distance (the distance among samples of a cluster), b is the distance between the sample and the closest cluster to which the sample does not belong. The best value and the worst value are 1 and -1, respectively, and values close to 0 indicate overlapping clusters.

DBI quantifies average similarity between each cluster and its most similar one using intra-cluster scatter and between-cluster separation:

$$DBI = \frac{1}{k} \sum_{i=1}^{k} \max_{i \neq j} \frac{s_i + s_j}{d_{ij}} \tag{12}$$

Where $s_i$ and $s_j$ represents intra-cluster distances for clusters $i$ and $j$, respectively; $d_{ij}$ represents the distance between cluster centroids $i$ and $j$.

CHI measures the ratio of between-cluster dispersion to within-cluster dispersion:

$$CHI = \frac{Tr(Wk)}{Tr(Bk)} \times \frac{N - K}{K - 1} \tag{13}$$

Where $Tr(Wk)$ represents a trace of the between-cluster dispersion matrix; $Tr(Bk)$ is the trace of the within-cluster dispersion matrix; N represents the number of data points; K is the number of clusters.

To assess the robustness of the clustering process, a sensitivity analysis was performed by varying the initial random seeds from 0 to 29 and repeating the clustering procedure for each candidate number of clusters. The mean and std of inertia, SC, DBI, and CHI across the ten runs were then computed to evaluate the stability of clustering outcomes.

### 2.5.2 Regression Modelling of PET

The modelling framework formulated both the mean of PET ($PET_{mean}$) and its standard deviation ($PET_{std}$) as functions of several remote sensing features and reference meteorological features, illustrated in Equation (14) and Equation (15), respectively.

$$PET_{mean} = f(T_{air_{ref}}, Solar_{ref}, V_{a_{ref}}, NDVI, MNDWI, NDBI, BSI, \alpha_{land}, \epsilon_{land}) \quad (14)$$

$$PET_{std} = f(T_{air_{ref}}, Solar_{ref}, V_{a_{ref}}, NDVI, MNDWI, NDBI, BSI, \alpha_{land}, \epsilon_{land}) \quad (15)$$

Where $T_{air_{ref}}$ is the reference dry-bulb air temperature; $Solar_{ref}$ is the reference global horizontal solar irradiance; $V_{a_{ref}}$ is the reference wind speed; $NDVI$ is the normalized difference vegetation index; $MNDWI$ is the modified normalized difference water index; $NDBI$ is the normalized difference built-up index; $BSI$ is the bare soil index; $BAEI$ is the built-up area extraction index; $\alpha_{land}$ is the land surface albedo; $\epsilon_{land}$ is the land surface emissivity.

In addition, as shown in Equation (16), $PET$ model based on deterministic method without considering the potential distribution was established using same input features for model performance comparison.

$$PET = f(T_{air_{ref}}, Solar_{ref}, V_{a_{ref}}, NDVI, MNDWI, NDBI, BSI, \alpha_{land}, \epsilon_{land}) \quad (16)$$

To predict these PET statistics, XGBoost (Chen et al., 2024) was employed with hyperparameters tuned across specific search spaces using Bayesian optimization from BayesSearchCV (Head et al., 2018), as shown in Table 3.

Table 3: Hyperparameter optimization range

| Hyperparameter | Range |
|---|---|
| n_estimators | (50, 500) |
| learning_rate | (0.01, 0.3) |
| max_depth | (2, 10) |
| subsample | (0.5, 1.0) |

Model performance was evaluated based on 5-fold cross-validation using Symmetric Mean Absolute Percentage Error (SMAPE), Root Mean Squared Error (RMSE) and the coefficient of determination ($R^2$). SMAPE penalizes the over-predictions and under-predictions symmetrically, suitable for actual values closing to zero, whereas RMSE penalizes larger errors in the residual squared process and is more interpretable in the same units as the original data.

$$\mathrm{SMAPE} = \frac{100}{\mathrm{n}} \sum_{i=1}^{n} \frac{2 \cdot \left|\mathrm{Y_i} - \widehat{\mathrm{Y}}_{\mathrm{i}}\right|}{\left|\mathrm{Y_i}\right| + \left|\widehat{\mathrm{Y}}_{\mathrm{i}}\right|} \quad (16)$$

$$\mathrm{RMSE} = \sqrt{\frac{1}{\mathrm{n}}\sum_{\mathrm{i}=1}^{\mathrm{n}}\left(\mathrm{Y_i} - \widehat{\mathrm{Y}}_{\mathrm{i}}\right)^{2}} \tag{17}$$

$$R^2 = 1 - \frac{\sum_{i=1}^{n}\left(Y_i - \widehat{Y}_i\right)^2}{\sum_{i=1}^{n}\left(Y_i - \overline{Y}\right)^2} \tag{18}$$

Where n represents the number of observations; $Y_i$ denotes a given observation; $\widehat{Y}_i$ expresses the predicted value.

Model interpretation was conducted based on SHapley Additive exPlanations (SHAP) using shap Python package (Lundberg and Lee, 2017). SHAP quantifies the contribution of each input feature to individual model prediction, providing an interpretable link between predictors derived from satellite indices (e.g., *NDVI*, *NDBI)* and environmental variables (e.g., $\mathrm{T_a}$, $\mathrm{V_a}$) and the predicted mean and standard deviation of PET. The summary plot of SHAP provides the overall importance and effect direction of key features, while dependence plots were created to illustrate the marginal effect of each feature on both mean and standard deviation of PET.

For PET inference, machine learning models were used to predict the $PET_{mean}$ and $PET_{std}$ based on reference meteorological data and remote sensing data at each grid point. Monte Carlo simulations with sampling size of 5000 were then implemented to draw repeated random samples from normal distribution determined by predicted $PET_{mean}$ and $PET_{std}$.

## 3. Results

### 3.1 Clustering of Reference Weather Data

As indicated in Figure 5, the clusters became more visually distinct and isolated as the number of clusters increased from 2 to 7. However, this reduced the data density within each cluster, thus limiting the amount of information available for model training. Consequently, while more clusters provided finer spatial detail, they may produce microclimate models with poor robustness or generalization ability due to the reduced representativeness of the data within each cluster. Figure 6 and Table 4 indicated that the inertia score at 3 clusters was 17.39±0.05, indicating a significant reduction from the inertia score of 2 clusters (26.07±0), while remaining relatively lower compared to higher cluster numbers, where adding more clusters did not substantially improve clustering quality. The SC at 3 clusters was 0.38±0, which was the second highest value compared to other clustering results. The DBI value at 3 clusters was 0.89±0.02, slightly higher than the lowest DBI value observed at 7 clusters (0.88±0.03). The CHI (312.89±1.52) was highest at 3 clusters indicating that the clusters were well separated and compact at this point. These performance metrics suggested that 3 clusters provided a good balance between cluster separation and compactness for reference meteorological data.

The dates associated with Cluster 1 (Figure 5) of the optimal 3 clusters were extracted to filter out the representative local microclimate data. Table 5 summarises the descriptive statistics of this reference weather subset, signifying a clear, hot, and rain-free ($Rain_{tot}$= 0 mm) weather condition. The daily mean of $T_{avg}$, $T_{max}$, $T_{min}$ are 29.09°C, 32.25°C, 26.84°C, respectively. The daily $Solar_{max}$ was 838.86±58.99 W/m$^2$, and the daily $Wind_{avg}$ was 3.27±0.61 m/s.

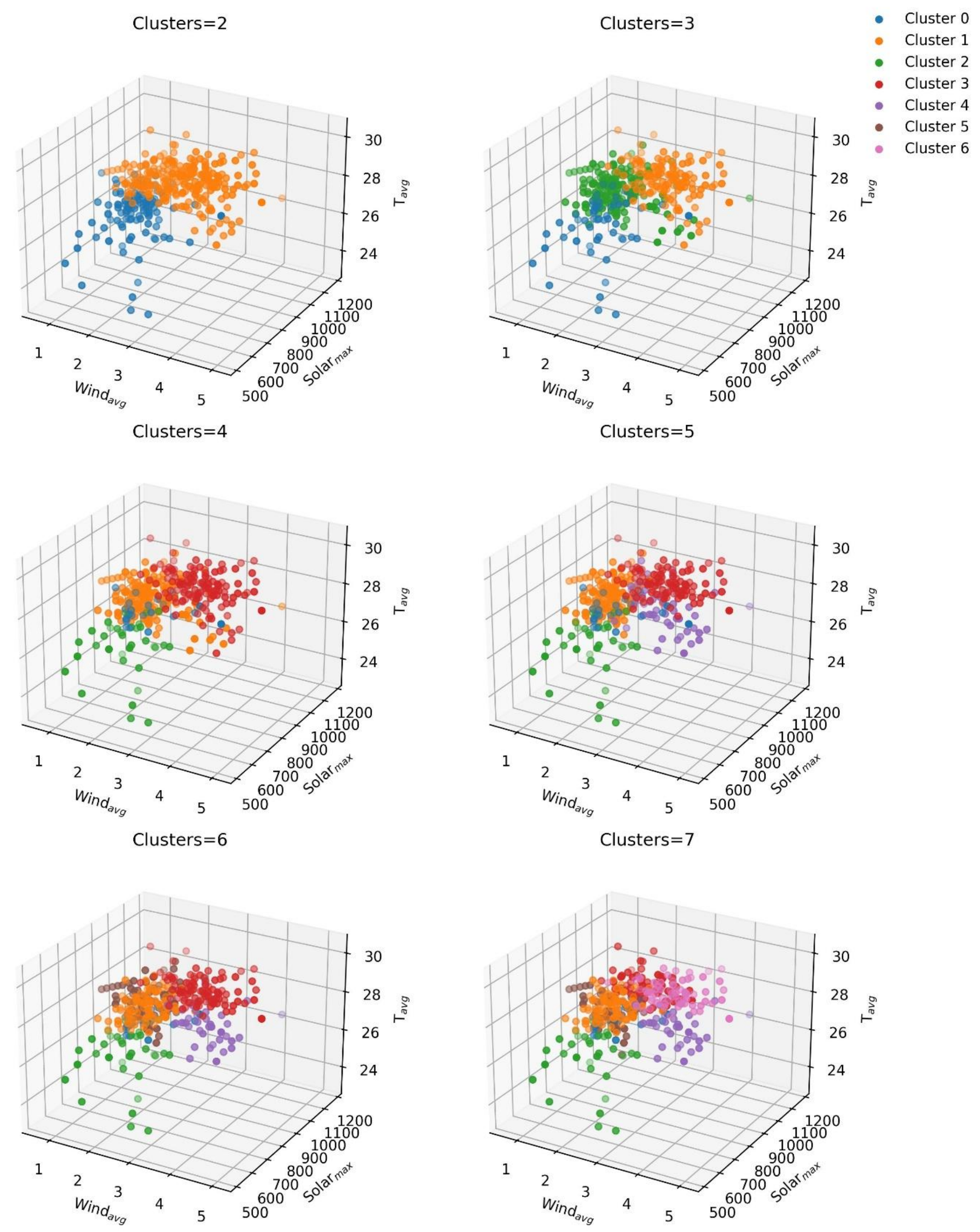

Figure 5: Clustering of reference weather data

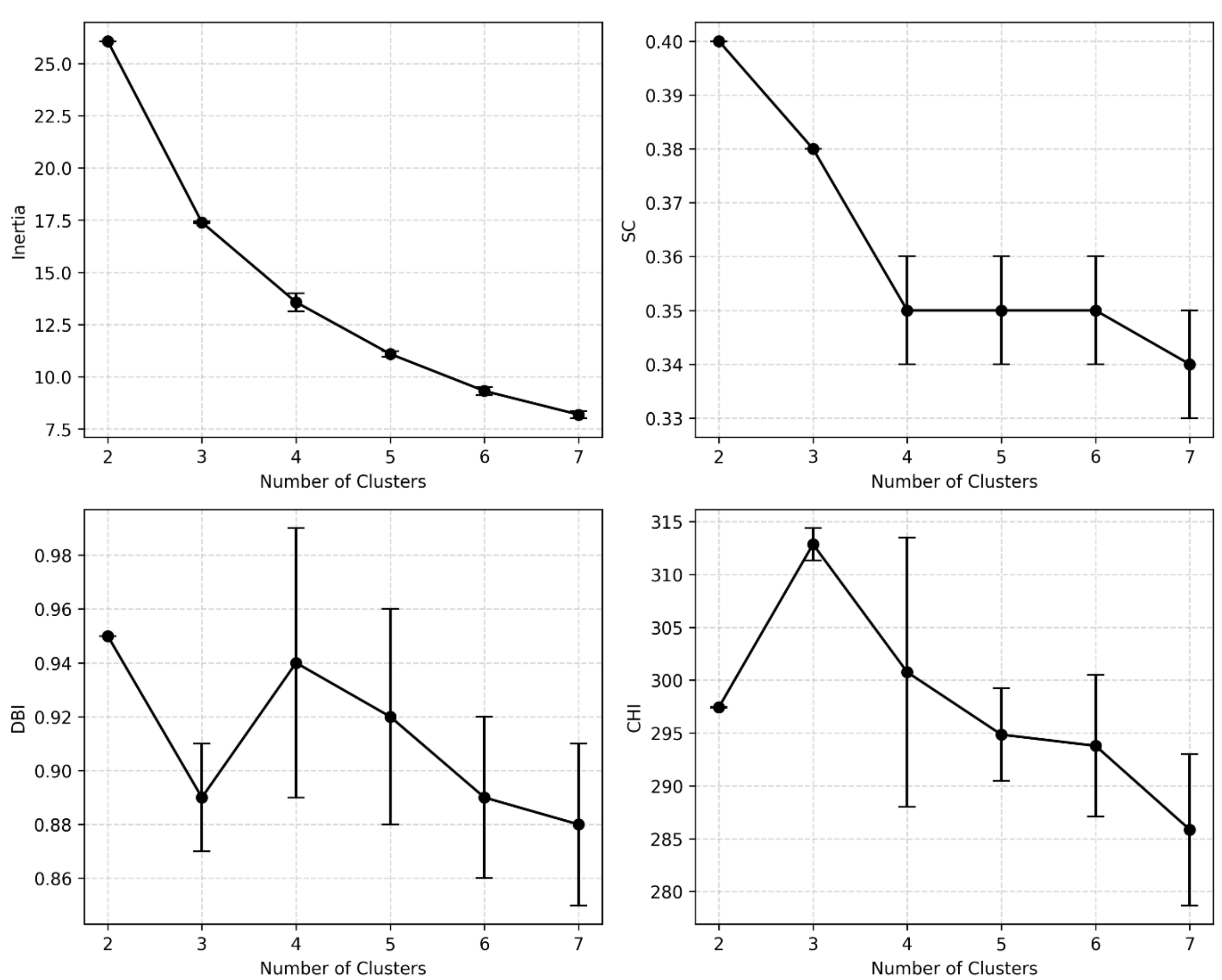

Figure 6: Clustering performance of reference weather data

Table 4: The sensitivity analysis of cluster numbers

| Number of Clusters | Inertia | | SC | | DBI | | CHI | |
|---|---|---|---|---|---|---|---|---|
| | Mean | Std | Mean | Std | Mean | Std | Mean | Std |
| 2 | 26.07 | 0 | 0.4 | 0 | 0.95 | 0 | 297.42 | 0.05 |
| 3 | 17.39 | 0.05 | 0.38 | 0 | 0.89 | 0.02 | 312.85 | 1.52 |
| 4 | 13.57 | 0.44 | 0.35 | 0.01 | 0.94 | 0.05 | 300.75 | 12.74 |
| 5 | 11.09 | 0.13 | 0.35 | 0.01 | 0.92 | 0.04 | 294.84 | 4.37 |
| 6 | 9.32 | 0.18 | 0.35 | 0.01 | 0.89 | 0.03 | 293.78 | 6.7 |
| 7 | 8.19 | 0.17 | 0.34 | 0.01 | 0.88 | 0.03 | 285.85 | 7.15 |

Table 5: Descriptive statistics of clustered weather data under clear and hot conditions

| Statistic | $Rain_{tot}$ (mm) | $Wind_{avg}$ (m/s) | $Solar_{tot}$ ($W/m^2$) | $Solar_{max}$ ($W/m^2$) | $Solar_{avg}$ ($W/m^2$) | $T_{avg}$ (°C) | $T_{max}$ (°C) | $T_{min}$ (°C) |
|---|---|---|---|---|---|---|---|---|
| mean | 0 | 3.27 | 7077.89 | 838.86 | 294.91 | 29.09 | 32.25 | 26.84 |
| min | 0 | 1.65 | 6117.5 | 740.4 | 254.9 | 27.26 | 30 | 25 |
| 25% | 0 | 2.87 | 6695.45 | 798.1 | 278.98 | 28.67 | 32 | 26 |
| 50% | 0 | 3.22 | 7017.4 | 825.15 | 292.39 | 29.1 | 32 | 27 |
| 75% | 0 | 3.72 | 7424.75 | 877.4 | 309.36 | 29.51 | 33 | 27.5 |
| max | 0 | 4.72 | 8747.1 | 1014.6 | 364.46 | 30.42 | 34.5 | 28.9 |
| std | 0 | 0.61 | 569.13 | 58.99 | 23.71 | 0.62 | 0.9 | 0.85 |

### 3.2 Spatial and Temporal Heterogeneity of PET Across Sampled Locations

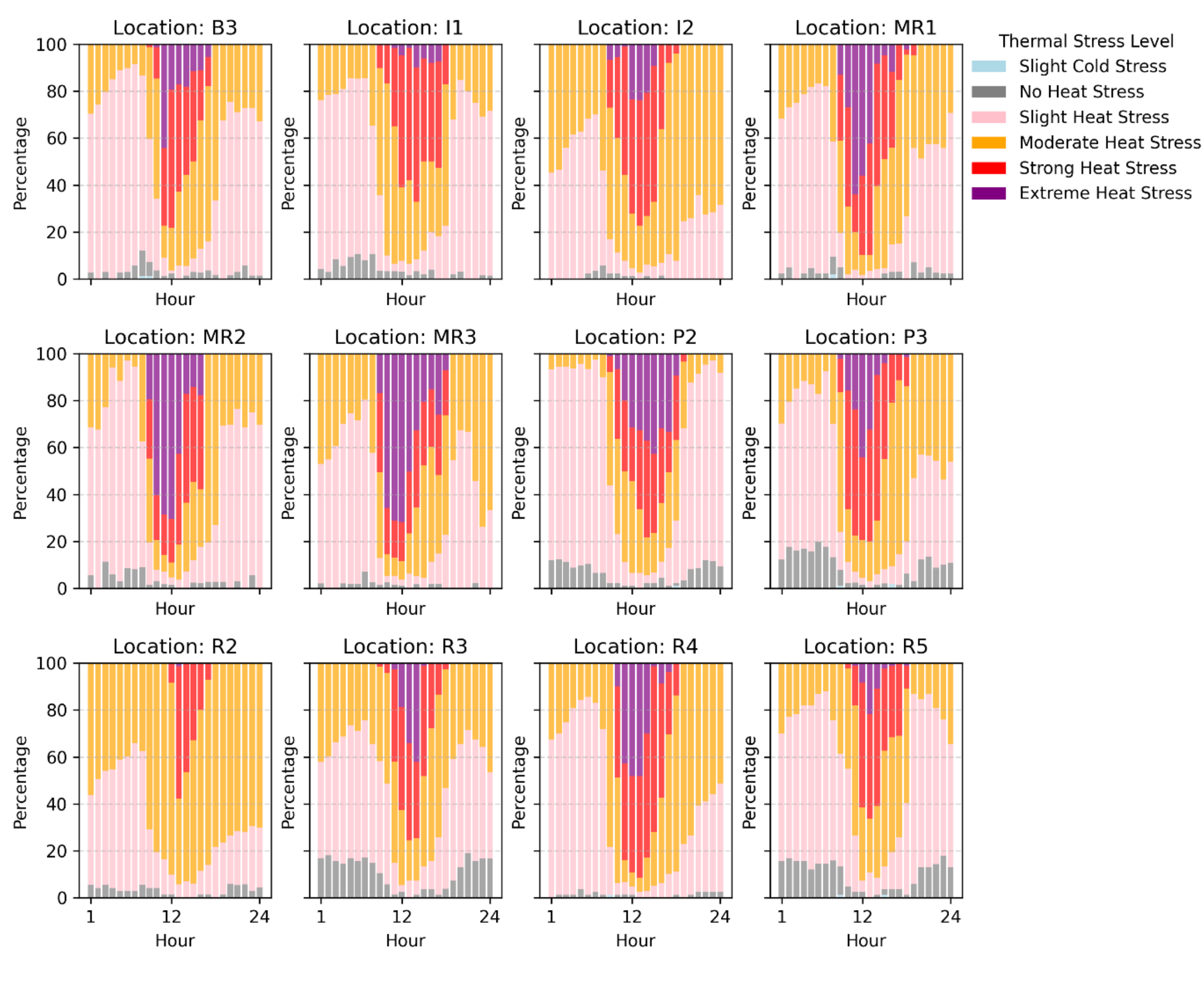

Figure 7: Spatial heterogeneity of PET across sampled locations

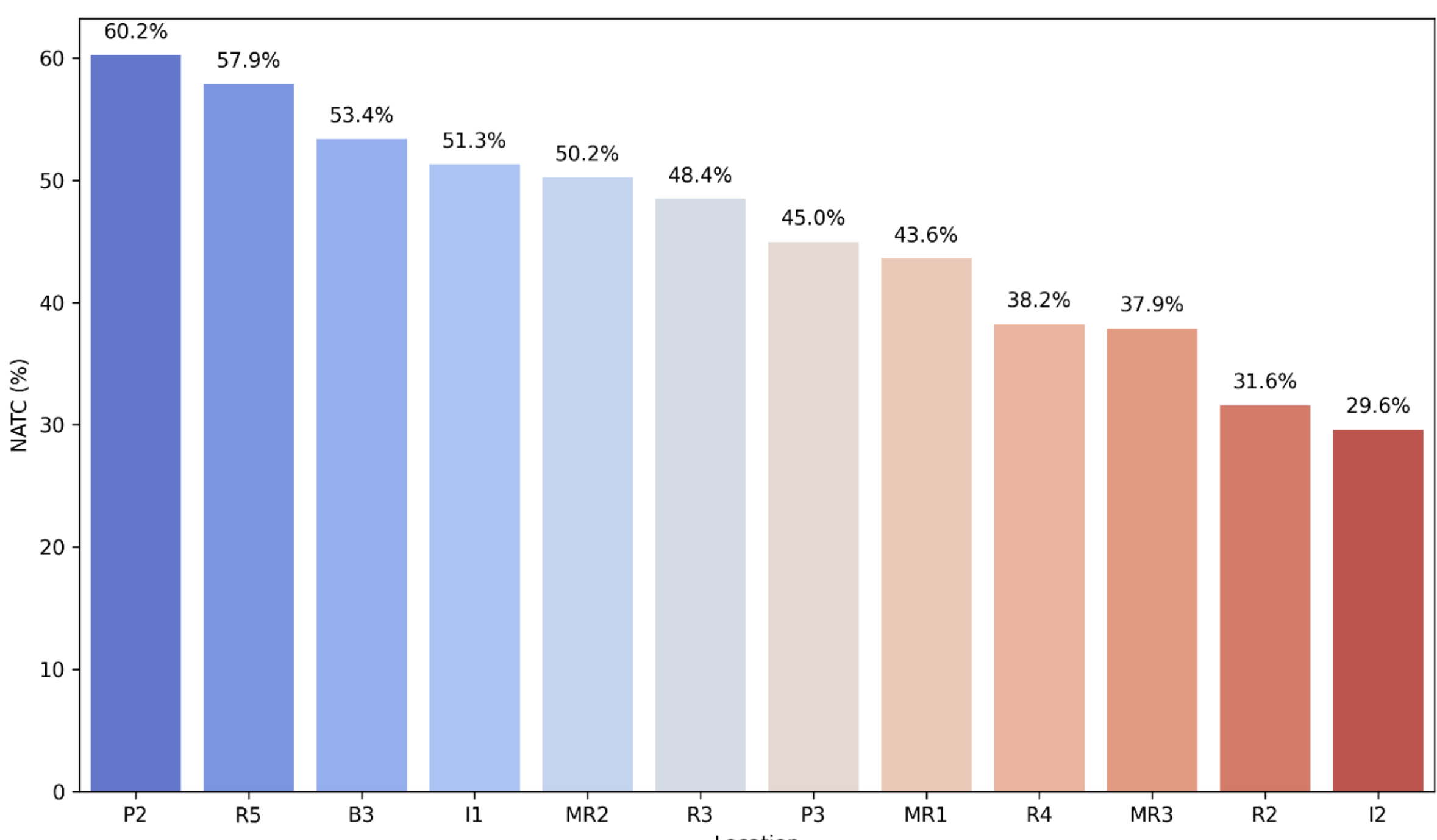

Figure 8: NATC across sampled locations

The thermal stress levels were categorized into six groups: slight cold stress, no heat stress, slight heat stress, moderate heat stress, strong heat stress, and extreme heat stress. Figure 7 summarizes the diurnal distribution of thermal stress levels at the 12 sampled locations. Across all locations, a clear diurnal pattern emerged. During the early morning and late evening hours, conditions were dominated by no and slight heat stress, occasionally accompanied by moderate heat stress. In contrast, moderate, strong and extreme heat stress were concentrated between late morning and mid-afternoon, coinciding with the period of peak solar loading. In addition to this common diurnal cycle, the panels revealed pronounced spatial differences in thermal exposure. MR2 and MR3 exhibited large percentages of moderate to strong heat stress over extended daytime periods, reflecting the combined effect of large areas of hard surfaces, limited shading and strong anthropogenic heat sources, while others might had more frequent occurrences of no heat stress (e.g., P2, P3, R3, R5). To summarize this spatiotemporal information into a single metric, the NATC was computed as the percentage of hours falling within acceptable categories over the 24-hour period at each site.

As illustrated in Figure 8, the park site (P2) consistently exhibited high percentages of no heat stress and slight heat stress throughout the day, resulting in the highest NATC (60.2%). This can be attributed to the park's vegetative cover reducing land surface temperature and mean radiant temperature through shading and evapotranspiration. Another park site P3 (NATC= 45%) without dense tree canopy suggested a warmer environment in the evening. In the commercial district, B3 (53.4%), located centrally among buildings, maintained relatively high thermal acceptability, consistent with the shading effects of the surrounding urban canyon that mitigated peak radiant heat. Along major roads, MR2 (50.2%) adjacent to P2 showed relatively higher thermal comfort level. While MR1 (43.6%) and MR3 (37.9%), both situated near built-up areas, showed comparable NATC values with significant moderate heat stress indicating severe discomfort conditions due to its open roadway setting dominated by asphalt and lacking significant shading. Among residential sites, R5 (57.9%) exhibited mixed but generally acceptable thermal conditions, whereas R2 (31.6%) had the lowest overall acceptable thermal comfort and the highest percentage of moderate heat stress over 24 hours, indicating a persistent high temperature environment. I2 is located in an industrial area with a large low-rise building area and no greenery cover. It has the worst thermal comfort with a NATC of 29.6%.

The hourly heat stress analysis provides detailed information on when and where thermal comfort issues occurred. However, it required further analysis such as regression models to find out why certain locations exhibit different thermal comfort conditions under different urban morphologies, which is crucial for heat mitigation strategies to improve thermal comfort in specific locations and at specific times.

### 3.3 Probabilistic Prediction of PET

$PET_{mean}$ was well-predicted by the optimized XGBoost model, achieving $R^2$, RMSE and SMAPE of 0.93, 0.81 ℃ and 1.34%, respectively (Table 6). In contrast, the prediction of $PET_{std}$ was more challenging with $R^2$, RMSE and SMAPE reaching 0.85, 0.38 ℃ and 10.44%, respectively. Meanwhile, $PET$ based on deterministic method had poorest performance with $R^2$ and RMSE of 0.73 and 2 ℃, respectively.

The distribution plots of $PET_{mean}$ and $PET_{std}$ at each location showed that the original values were in good agreement with the predicted values (Figure 9). The predicted values of some locations (e.g., B3, I1, I2, MR1-MR3, P2, and R2) were lower than the original peak values. The residual analysis indicated that for $PET_{mean}$, most residuals were tightly clustered around 0 °C, with the majority falling within ± 2 °C and only a few observations exceeding 4 °C (Figure 9). The residuals were randomly distributed without temporal trends, although a slight underestimation was observed at higher PET values. For $PET_{std}$, residuals were narrower in absolute magnitude within ± 1 °C but exhibited a systematic pattern: higher observed variability tended to be underestimated, while lower variability was slightly overestimated, reflecting a regression-to-the-mean effect.

Table 6: Bayesian optimization results of hyperparameters and model performance of XGBoost

| Target | learning_rate | max_depth | n_estimators | subsample | $R^2$ | RMSE (℃) | SMAPE (%) |
|---|---|---|---|---|---|---|---|
| $PET_{mean}$ | 0.16 | 7 | 328 | 1 | 0.93 | 0.81 | 1.34 |
| $PET_{std}$ | 0.06 | 9 | 273 | 0.92 | 0.85 | 0.38 | 10.44 |
| $PET$ | 0.07 | 7 | 500 | 1 | 0.73 | 2.00 | 3.94 |

a. PET distribution at different locations

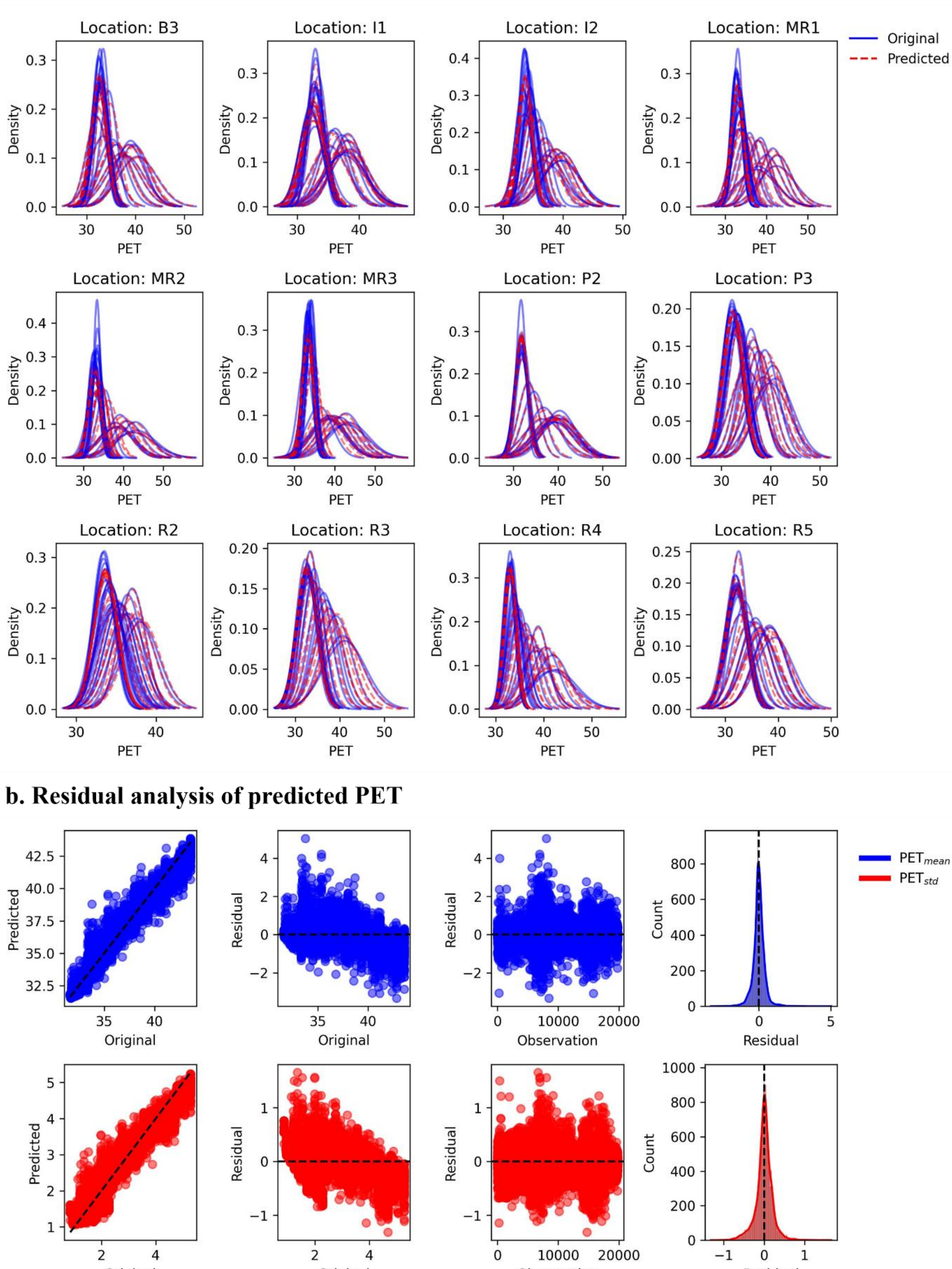

Figure 9: Evaluation of regression models: (a) Original and predicted PET distribution at different locations; (b) Residual analysis of mean and standard deviation of PET

Figure 10 presents the importance of input features and the partial dependence between various urban morphological features and their corresponding SHAP values for $PET_{mean}$ and $PET_{std}$. The SHAP summary plot reveals that *NDVI* was the most important urban morphological predictor, followed by *MNDWI*, $\alpha_{land}$, *NDBI*, *BSI* and $\epsilon_{land}$. For $PET_{mean}$, decreasing SHAP values were associated with increasing *MNDWI* reflecting the well-established cooling influence of water bodies through evaporative advection. In contrast, the *NDBI* displayed a positive relationship with increasing SHAP values. This reflects the urban heat island effect from built-up areas. *NDVI* demonstrated a more complex and nonlinear relationship where both cooling and warming effects emerged. Lower NDVI had a warming effect on $PET_{mean}$, whereas NDVI > 0.2 had a cooling effect on $PET_{mean}$, indicating that the cooling potential of green spaces is related to the health of plants. $\alpha_{land}$ also exhibited a complex cooling potential but is generally negatively correlated with PET. In terms of $PET_{std}$, both *NDVI* and *MNDWI* were nonlinearly correlated with SHAP values indicating complex effects of vegetation cover and water bodies on thermal comfort variability. Meanwhile, the *NDBI* was positively correlated with increasing SHAP values indicating that thermal comfort variance was higher in areas with higher built-up areas and bare soil. Regarding $\alpha_{land}$, it consistently exerted a negative influence indicating that reflective surfaces reduced both mean heat load and short-term variability by lowering daytime heat gain. $\epsilon_{land}$ and *BSI* showed negligible influence for both $PET_{mean}$ and $PET_{std}$.

Figure 11 shows the interaction effects of key predictors on PET. For $PET_{mean}$, lower NDVI combined with higher NDBI produced positive SHAP values, indicating greater thermal stress in densely built-up areas. As NDVI increased, MNDWI values varied widely, and both high and low water presence were observed across vegetation gradients. When NDVI was below 0.2, higher $\alpha_{land}$ resulted in negative SHAP values, whereas when NDVI exceeded 0.2, higher albedo shifted SHAP values to the positive range. NDBI consistently contributed positive SHAP values, but this effect weakened under higher MNDWI. High NDBI coupled with high albedo increased SHAP values, while high albedo near water bodies produced negative SHAP values. For $PET_{std}$, lower NDVI combined with higher NDBI led to lower variability compared to areas with higher NDVI and lower NDBI. Lower NDVI with higher MNDWI also resulted in reduced $PET_{std}$, whereas increases in NDVI reduced $PET_{std}$ slightly when MNDWI decreased. Lower $PET_{std}$ was further associated with high $\alpha_{land}$ in areas with low NDVI. The effects of NDBI with MNDWI and NDBI with albedo were marginal. In contrast, higher MNDWI together with higher $\alpha_{land}$ led to increased $PET_{std}$.

**a. Feature importance**

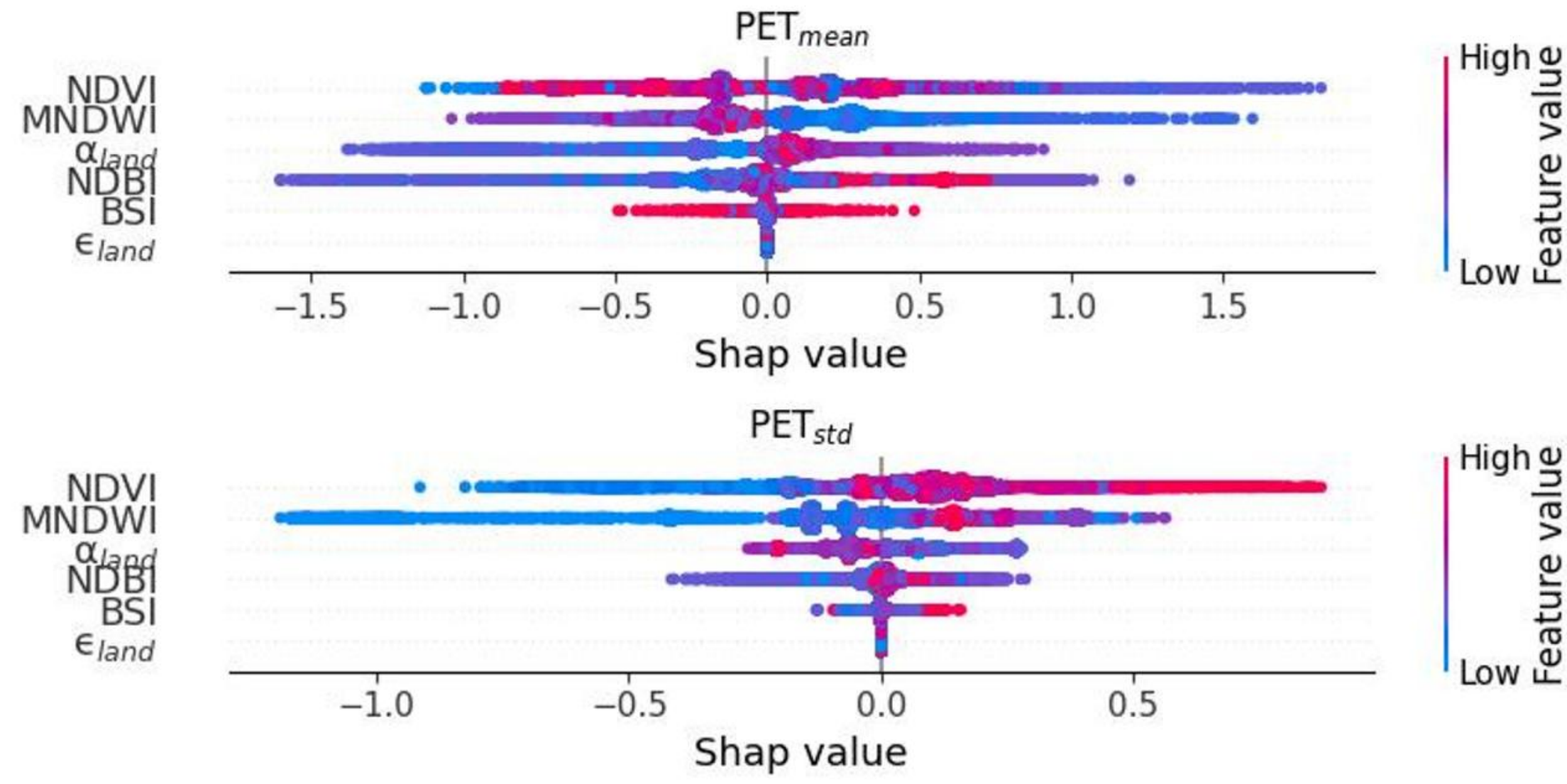

**b. Dependence analysis**

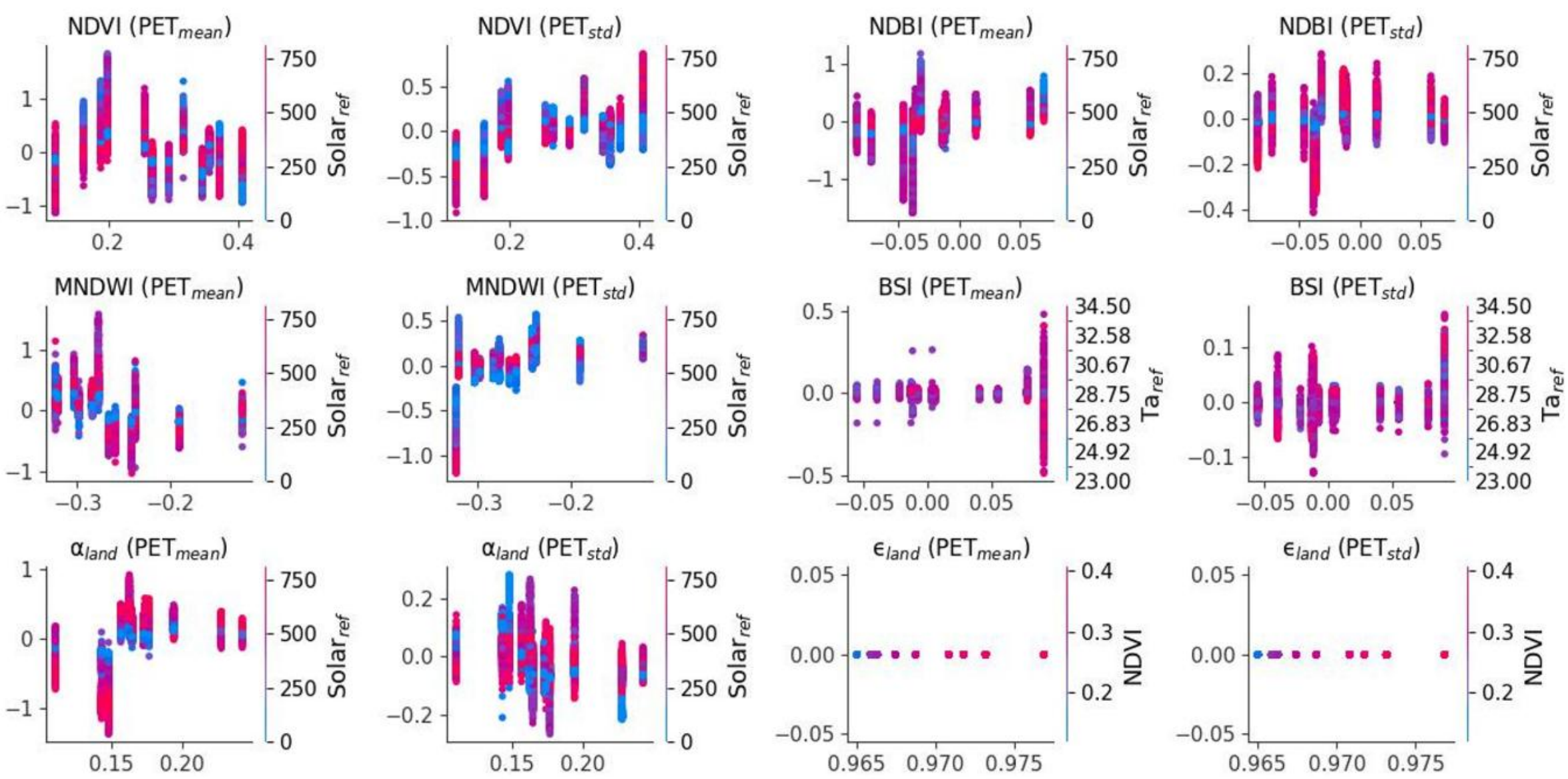

Figure 10: SHAP analysis of $PET_{mean}$ and $PET_{std}$: (a) Feature importance; (b) Dependence analysis of individual feature

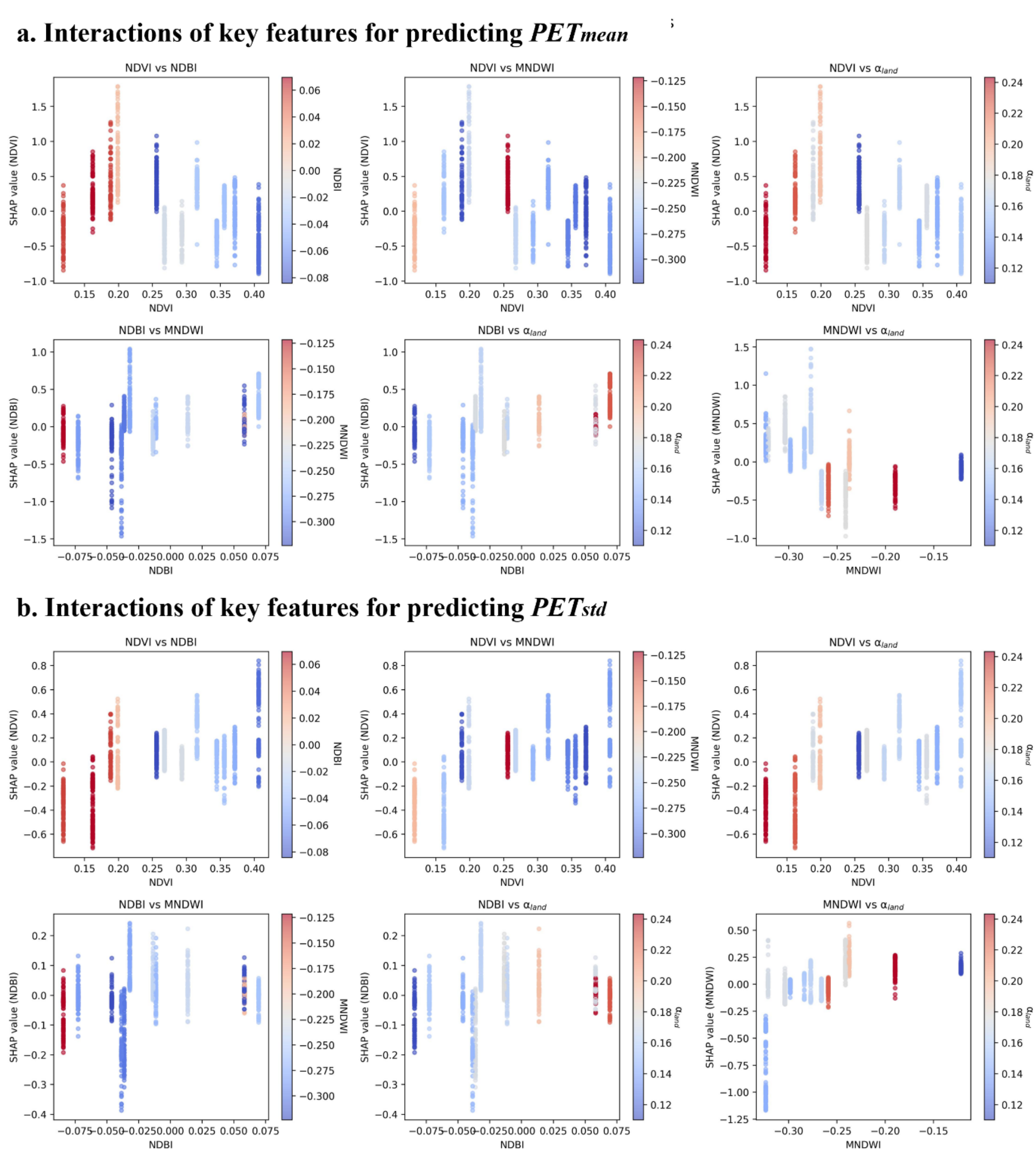

Figure 11: SHAP analysis of interaction effects among key features: (a) $PET_{mean}$; (b) $PET_{std}$

Outdoor thermal comfort assessments were conducted in the Jurong East area using trained models of $PET_{mean}$ and $PET_{std}$. The prediction of NATC at Jurong East area (Figure 12) reveals a pronounced spatial heterogeneity in outdoor thermal comfort. Comfortable areas with higher NATC values are typically located near parks, gardens, or other green spaces. Conversely, uncomfortable areas with lower NATC values are often associated with urban development and infrastructure. Locations near to greenery cover, particularly areas with dense tree canopies and vegetated surfaces, achieve NATC up to 65%. This is primarily due to the shading effect provided by vegetation reducing mean radiant temperature by blocking the direct solar radiation. In contrast, areas dominated by built-up surfaces, especially those with extensive impervious cover such as industrial or high-density residential areas, have NATC values below 30%. Due to the urban heat island effect, these areas have higher temperatures, and materials such as concrete and asphalt absorb and retain heat while releasing longwave radiation back to pedestrians. For $PET_{mean}$ (Figure 13), the heatmap revealed a clear contrast between daytime (08:00–19:00) and nighttime (20:00–07:00) thermal conditions. The maximum $PET_{mean}$ difference between hot and cold spots reached approximately 1.5 °C and 6 °C for nighttime and daytime, respectively. This indicated a more pronounced thermal comfort difference during the day compared to the night. At night, cold spots were prominent in areas with dense greenery, while hot spots were concentrated in heavily built-up zones, particularly industrial areas. During the day, areas with extensive tree canopy consistently exhibited the lowest PET, whereas some built-up areas also showed reduced PET due to self-shading effects. In terms of $PET_{std}$ (Figure 14), greenery covered areas consistently showed lower thermal variability compared to built-up areas. During the daytime (08:00–19:00), thermal variability ranged from 4.0–4.5 °C in built-up areas and 1.5–2.0 °C in greenery covered areas. At night (20:00–07:00), the variability was reduced to 2.2–2.4 °C in built-up areas and 1.2–1.4 °C in greenery covered areas. The consistently low temperature fluctuations during the day and night in green spaces suggested greenery played a key role in regulating temperature fluctuations and improving thermal comfort in urban environments.

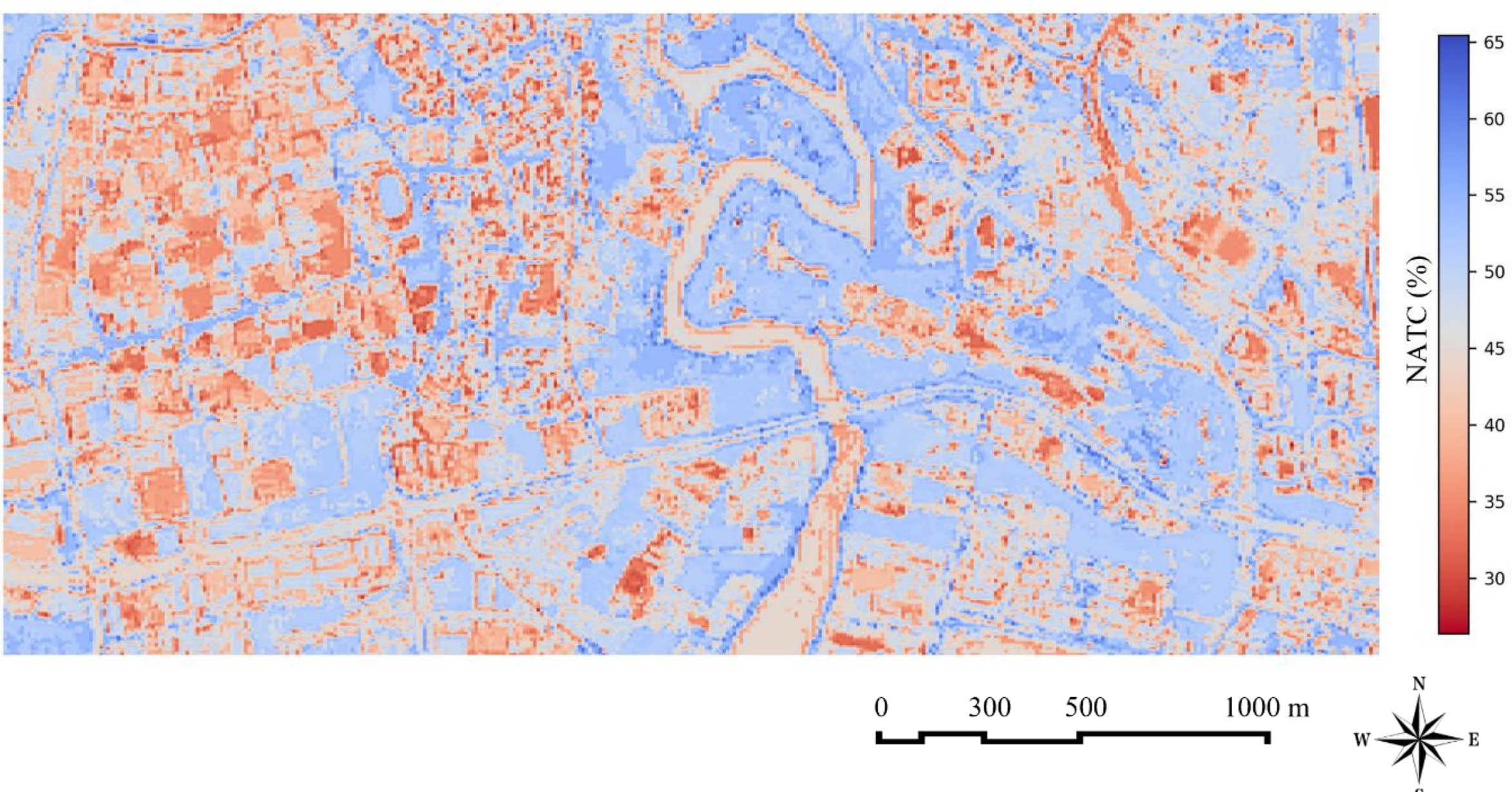

Figure 12: The prediction of NATC at Jurong East area

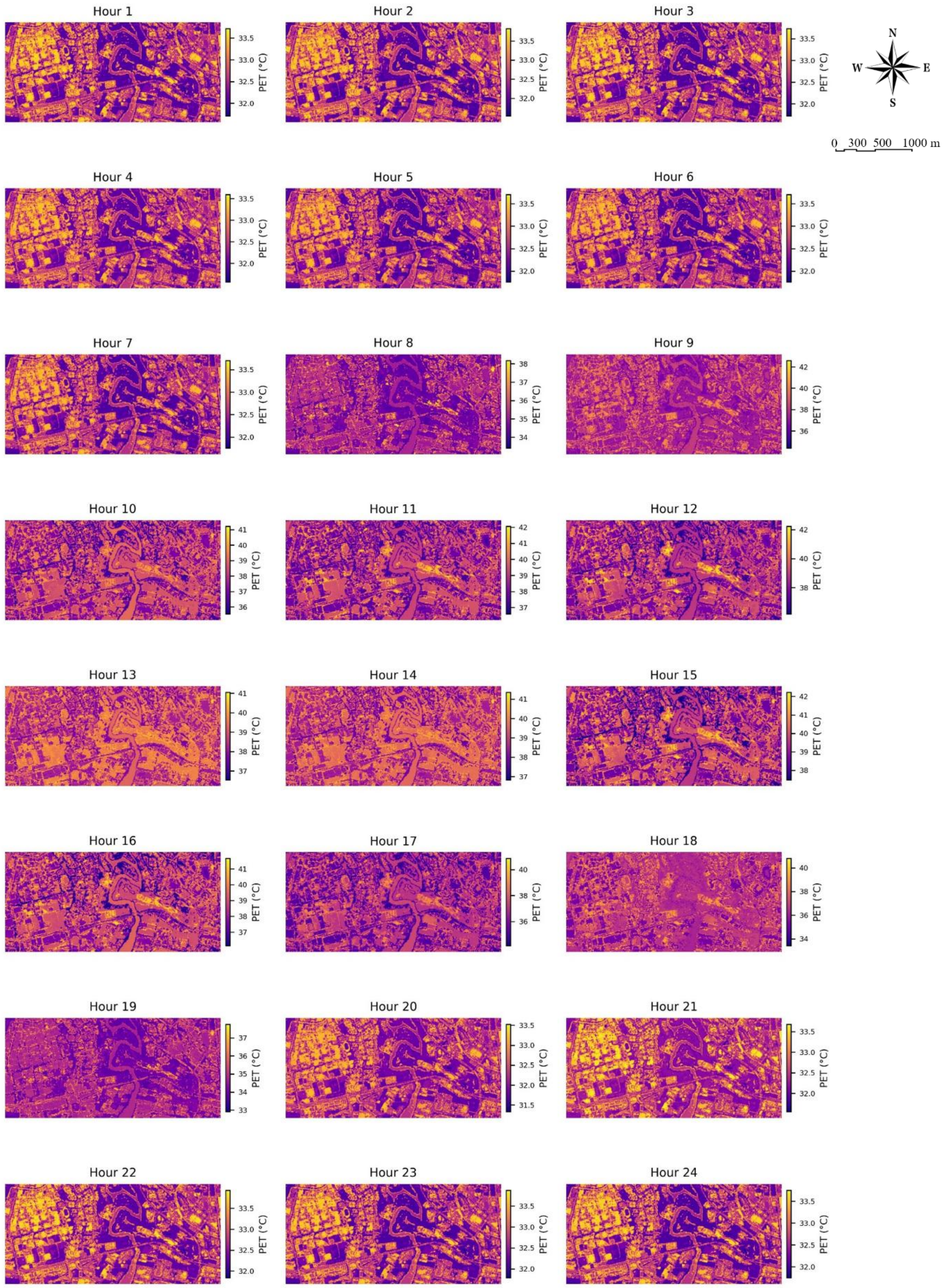

Figure 13: Hourly heatmap of $PET_{mean}$ at Jurong East area

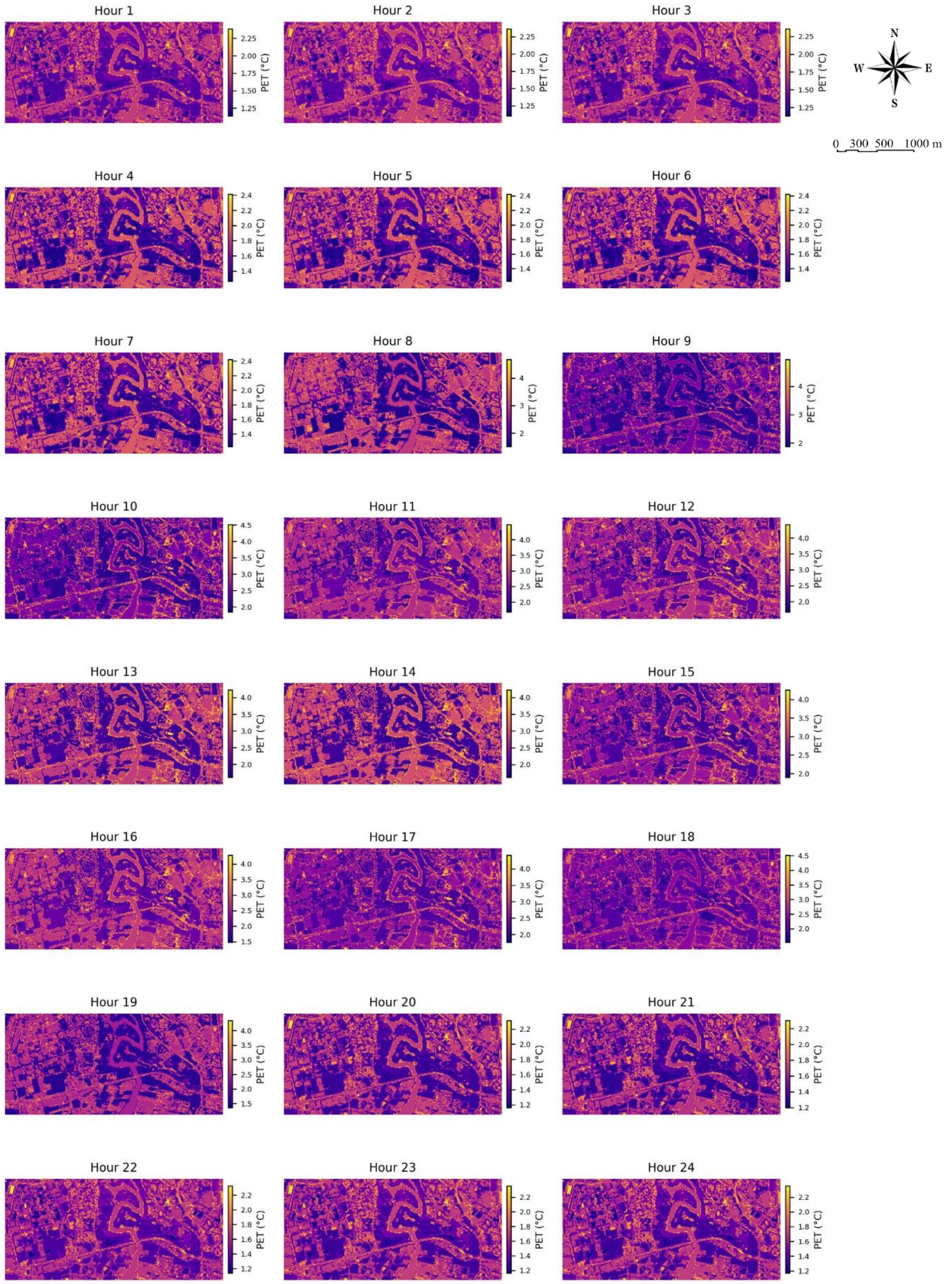

Figure 14: Hourly heatmap of $PET_{std}$ at Jurong East area

## 4. Discussion

### 4.1 Stochastic Representation of Outdoor Thermal Comfort

As shown in Figure 8, the underlying distribution of PET over longer time periods exhibited substantial variability under clear, hot and rain-free weather conditions driven by both microclimatic fluctuations and interactions with urban morphologies. Previous studies based on short-term measurements (Elgheznawy and Eltarabily, 2021; He et al., 2023; Miao et al., 2023; Ou and Lin, 2023; Sun et al., 2022; Tabatabaei and Fayaz, 2023; Zhao et al., 2023) often fail to capture the full range of thermal conditions. Furthermore, conventional regression modelling typically establish deterministic relations between a single thermal comfort index (e.g., PET, UTCI) and environmental or morphological indicators yielding average predictions but overlooking the temporal variability and probability of extreme stress levels (Anders et al., 2025; Deng et al., 2023; Jović et al., 2016; Pantavou et al., 2022; Young et al., 2022; Zhang et al., 2022).

In terms of numerical simulation, urban microclimate simulators (e.g., ENVI-met v5.8.0) solve the coupled momentum, heat and radiation balances to produce gridded fields of air temperature, wind speed, humidity and radiative fluxes at metre scale resolution. After a simulation is completed, the BIO-met module acts as a post-processor that computes biometeorological indices for user-specified human parameters (e.g., posture/activity, metabolic rate, clothing insulation) at each grid cell and time step. While this workflow is well suited to scenario analysis, it is inherently deterministic at run time. As reported in previous studies (Schöneberger et al., 2025; Zhang et al., 2025), for a given geometry, material parameterisation, boundary conditions and human settings, the model returns a single trajectory of microclimate and a single index value per time step, resulting in average conditions. This imposes several limitations on thermal comfort assessments: input uncertainties are not propagated to PET prediction intervals, BIO-met can only produce point estimates, and high computational costs constrain ensemble simulations limiting the evaluation of design options.

In contrast, a stochastic approach explicitly represents thermal comfort as a distribution rather than a fixed value. By repeatedly sampling from the predicted PET distribution, stochastic simulations can capture the range of possible PET outcomes. This not only allows for an estimate of central tendency but also quantifies the uncertainty and risk associated with extreme heat stress. Since thermal comfort perception is probabilistic in nature caused by microclimate fluctuations, human metabolism, and individual clothing levels (Tartarini and Schiavon, 2020), it is essential to estimate both the expected conditions and the uncertainty around them. Human metabolism and personal clothing levels do not vary much within a given season. For example, people tend to walk and wear typical summer clothing during summer activities, therefore human metabolism and personal clothing levels were assigned to be 2 met and 0.5 clo in this study, respectively. In contrast, microclimate fluctuations strongly depend both on macroclimatic variations and on the surrounding urban morphologies. The latter is due to the fact that the urban morphologies have significant impacts on urban climate by altering the heat balance and wind characteristics within the urban canopy (Chen et al., 2020; Peng et al., 2022; Tong et al., 2017; Wu et al., 2022; Yu et al., 2020; Yuan et al., 2020), where a stochastic framework can provide a more realistic description of local variability and heat exposure.

The proposed framework advances urban thermal environment assessment by employing stochastic simulations characterizing heat stress events as a probabilistic distribution rather than single outcome. This approach not only quantifies average thermal comfort but also reveals the intensity and likelihood of extreme heat stress events that deterministic model tends to overlook. This enables more realistic assessments of heat exposure reflecting the heterogeneous microclimate dynamics within urban areas. It is also crucial for climate responsive design as adaptation strategies must address a range of plausible scenarios rather than optimizing for average conditions.

### 4.2 Key Drivers of Spatial and Temporal Thermal Stress

In terms of supervised machine learning, XGBoost proved effective for modelling nonlinear relationships between urban morphology and PET distributions achieving high predictive accuracy for $PET_{mean}$ ($R^2 = 0.93$). This is expected as nonlinear relationships have been discovered between urban morphologies and local urban climatic variables such as $T_a$ and $T_{mrt}$ (Chen et al., 2023; Gu et al., 2024; Jia et al., 2024; Liu et al., 2024; Wu et al., 2022). Predictions of $PET_{std}$ ($R^2 = 0.85$) were less precise underscoring the challenge of capturing fluctuations in thermal comfort. The lower predictive accuracy for hourly $PET_{std}$ was likely attributed from the complex and stochastic nature of short-term thermal variability, and the limited variance-specific information available in the model inputs, rather than limitations of XGBoost. Improving $PET_{std}$ predictions require additional variability focused predictors or feature engineering designed to better capture the environmental fluctuations. For example, incorporating hourly maximum and minimum values of key meteorological variables can provide models with enhanced information about short-term variability. The deterministic method performed poorly primarily because it attempted to reproduce PET at every single point in space and time, where high frequency variations and measurement uncertainties can amplify prediction errors. In contrast, representing thermal conditions using statistics such as the mean and standard deviation offered a more robust and realistic depiction of heat stress, reflecting both central tendency and variability. This probabilistic representation is more consistent with the inherent nature of microclimate condition and human thermal perception, where local fluctuations are less important than overall thermal exposure patterns.

For the computation of urban morphological indictors, multispectral remote sensing data can provide continuous spatial assessment of urban morphology data with fine-resolution (e.g., 10 m) and efficient processing, which has been mainly used to estimate the land surface temperature (LST) as the two data have similar scales and resolutions (Chen et al., 2006; Jamei et al., 2022; Jia and Wang, 2020; Mangiameli et al., 2022). However, it is not sufficient to use only LST to estimate PET or UTCI, which requires $T_a$, $T_{mrt}$, RH, and $V_a$ as environmental inputs. Therefore, ground-based microclimate measurements to capture the change of $T_a$, $T_{mrt}$, RH, and $V_a$ remain essential. Nonetheless, multispectral remote sensing data can help reduce reliance on 3D models, which are difficult to validate against the real world and often lack critical information such as building height at many sites. Combined with ground-based microclimate data, multispectral data represents a practical approach for estimating thermal comfort at the city scale.

This study identified *NDVI, MNDWI, NDBI*, and $\alpha_{land}$ as key drivers with significant importance in predicting outdoor thermal comfort. High albedo surfaces generally shift SHAP values into the negative range reflecting their ability to reflect a larger fraction of incident radiation and thus mitigate heat gain indicating a cooling influence (Lopez-Cabeza et al., 2022). However, their performance within the urban canopy appeared more complex, as reflected radiation can be trapped or reabsorbed by surrounding structures and vegetation. Whereas low albedo surfaces produced positive SHAP values, indicating greater heat absorption by hard materials with high thermal mass and specific heat capacity. Higher *MNDWI*, indicative of proximity to open water, is associated with negative SHAP values due to evaporative cooling through advection from water bodies (Liu et al., 2024). Vegetation represented by *NDVI* is generally expected to reduce PET through shading and evapotranspiration (Imran et al., 2021; Pérez et al., 2022). Higher *NDBI* values correspond to positive SHAP contributions reflecting the thermal amplification from dense built-up surfaces. While the factors influencing $PET_{mean}$ largely determine its absolute thermal conditions, the drivers of $PET_{std}$ shape its temporal variability reflecting the magnitude of intra-period (e.g., diurnal) fluctuations in thermal comfort, underscoring the interplay between solar radiation, surface moisture, and land-cover characteristics in modulating the amplitude of thermal comfort variability in tropical urban environments.

The interaction patterns showed that vegetation, built-up surfaces, water bodies, and albedo jointly shaped both mean thermal conditions and their variability. Low NDVI with high NDBI, together described sparsely vegetated and densely built-up surfaces, increased $PET_{mean}$ due to limited shading and evapotranspiration, whereas vegetation and water reduced $PET_{mean}$ through canopy shading and evaporative effects. Albedo had contrasting effects as it improved thermal comfort in open areas by reflecting most of the shortwave radiation directly back into the sky without being blocked by the tree canopy. However, it increased $PET_{mean}$ in dense vegetation where reflected radiation became trapped. The reduced influence of NDBI near water indicated that river breeze could moderate heat. High NDBI with high albedo intensified heat as reflected radiation being trapped and absorbed by the hard materials within the urban canopy. In contrast, high albedo near water enhanced cooling. For $PET_{std}$, low NDVI with high NDBI or high MNDWI produced lower variability because built-up and water-dominated areas were more thermally stable. Certain higher NDVI and lower NDBI conditions were associated with positive SHAP values for $PET_{std}$. This likely reflected the strong diurnal contrast created by tree shading, with substantial cooling in the morning and evening and intermittent temperature rise due to radiation around midday, resulting in larger fluctuations. High albedo in low vegetation areas further reduced variability, whereas high albedo with high MNDWI increased $PET_{std}$ by creating strong thermal contrasts.

Overall, this study revealed that both $PET_{mean}$ and $PET_{std}$ were lower in areas with healthy vegetation (NDVI >0.2) and relatively limited building cover. Previous studies have demonstrated that greenery such as tree canopy plays a critical role by providing shade, moderating temperatures (e.g., $T_{mrt}$, $T_a$), and enhancing evapotranspiration, thereby helping to reduce heat exposure and improve thermal stability in urban outdoor environments (Chen et al., 2023; Imran et al., 2021; Jia et al., 2024; Pérez et al., 2022). This study shows greenery can not only mitigate the intensity of heat stress by lowering average thermal conditions, but also reduce the variation of heat stress across the space resulting in a more stable and comfortable microclimate.

### 4.3 Limitations and Future Studies

As a single case study, the specific findings of this study have limited generalizability. However, the proposed framework is highly replicable which can be extended to other geographical locations and climate types with appropriate validation and testing. While the framework is adaptable to diverse regions and climatic conditions, its successful application requires the availability of ground-based meteorological observations and the retraining of machine learning models to account for local climatic and environmental characteristics. The 5-fold cross validation employed in this study was based on random sampling of the dataset, and the remote sensing index was extracted with a buffer size of 100 m radius, therefore the model may perform differently under different random states and buffer sizes. Expanding the dataset to include longer time spans, more diverse locations and different buffer sizes would enhance the model's robustness and applicability. In addition, the use of remote sensing data introduces uncertainties related to data resolution, cloud cover, and calibration. Future work should explore higher-resolution satellite data and improved calibration techniques to enhance the reliability of input data. This would also involve validating the model against additional datasets from other regions to ensure its accuracy and generalizability.

## 5. Conclusions

Microclimatic conditions in the urban canopy are influenced by variable radiation fluxes and complex wind fields, causing significant spatiotemporal variability in thermal comfort. This variability, due to surface-atmosphere energy exchange, air turbulence, and shading, shall not be modeled using deterministic methods. To address this gap, this study established a probabilistic framework for predicting the spatial and temporal dimensions of outdoor thermal comfort intensity and its variability, uncovering the stochastic nature of thermal comfort in different urban morphologies.

The framework demonstrated strong predictive performance, with $R^2 = 0.93$, RMSE = 0.81 °C, and SMAPE = 1.34% for $PET_{mean}$, and $R^2 = 0.85$, RMSE = 0.38 °C, and SMAPE = 10.44% for $PET_{std}$. Among the morphological predictors, *NDVI, MNDWI, NDBI*, and $\alpha_{land}$ emerged as key morphological drivers of spatiotemporal thermal stress, with *NDVI* being the dominant predictor.

In terms of case study assessment, locations with healthy vegetation such as extensive tree canopy and turf cover, recorded NATC values as high as 65%. In contrast, locations dominated by impervious built-up surfaces such as industrial estates and high-density residential blocks showed NATC values below 30%. Meanwhile, greenery plays a dual role in reducing the overall intensity of heat stress by lowering average thermal conditions and minimizing spatial variation in heat stress, thereby producing a more stable and comfortable microclimate. Specifically, daytime (08:00–19:00) PET variations ranged from 4.0–4.5 °C in built-up areas and 1.5–2.0 °C in vegetated areas, while nighttime (20:00–07:00) variations decreased to 2.2–2.4 °C in built-up areas and 1.2–1.4 °C in green cover areas.

These findings underscore the stochastic nature of thermal comfort across different urban morphologies and reveal how vegetation and built surfaces modulate not only average conditions but also the variability of heat stress. The proposed framework provides a practical, distribution-aware tool for accurate thermal comfort assessment and probability-informed planning and operations, particularly where detailed three-dimensional information is scarce, but satellite imagery and sparse ground observations are available. By estimating the likelihood of extreme heat stress and mapping risk, it enables the prioritization of interventions in high-risk regions.

**Acknowledgements**

This research was supported by: (1) The Fujian Province Young and Middle-aged Teacher Education and Research Project Funding (JAT241009); (2) Fuzhou University Research Starting Fund (511470); (3) Supporting Cooling NUS with Baseline-Evaluating-Action-Monitoring (BEAM) initiative from National University of Singapore.

**Conflicts of Interest**: The authors declare no conflict of interest.